\RequirePackage{lineno}
\documentclass[onecolumn,superscriptaddress,preprintnumbers,amsmath,amssymb,prc]{revtex4}  % double column
\usepackage{graphicx}% Include figure files
\usepackage{dcolumn}% Align table columns on decimal point
\usepackage{bm}% bold math
\usepackage{float}
\usepackage{color}
\usepackage{CJK}% Chinese character support
\usepackage[colorlinks,linkcolor=blue,urlcolor=blue,citecolor=blue]{hyperref}
\usepackage{isotope}
\usepackage[normalem]{ulem}
\renewcommand{\sout}{\bgroup \color{red} \ULdepth=-0.5ex \ULset}

\usepackage{amsmath}
\usepackage{cases}%the environment for piecewise functions
\usepackage{amssymb}
\usepackage{eqnalign}

\begin{document}
\begin{CJK*}{UTF8}{gbsn}% Use default fonts from CJK (see below)

\title{Constraining the chiral magnetic effect using spectator and participant planes across Au+Au and isobar collisions at $\sqrt{s_{_{\rm NN}}} = 200$ GeV}

\author{Bang-Xiang Chen}
\affiliation{Key Laboratory of Nuclear Physics and Ion-beam Application~(MOE), Institute of Modern Physics, Fudan University, Shanghai $200433$, China}
\affiliation{Shanghai Research Center for Theoretical Nuclear Physics, NSFC and Fudan University, Shanghai $200438$, China}

\author{Xin-Li Zhao}
\email{zhaoxinli@usst.edu.cn}
\affiliation{College of Science, University of Shanghai for Science and Technology, Shanghai $200093$, China}
\affiliation{Shanghai Research Center for Theoretical Nuclear Physics, NSFC and Fudan University, Shanghai $200438$, China}

\author{Guo-Liang Ma}
\email{glma@fudan.edu.cn}
\affiliation{Key Laboratory of Nuclear Physics and Ion-beam Application~(MOE), Institute of Modern Physics, Fudan University, Shanghai $200433$, China}
\affiliation{Shanghai Research Center for Theoretical Nuclear Physics, NSFC and Fudan University, Shanghai $200438$, China}

%\date{\today}% It is always \today, today,
             %  but any date may be explicitly specified

\begin{abstract}

We investigate the chiral magnetic effect (CME) in relativistic heavy-ion collisions through an improved two-plane method analysis of the $\Delta\gamma$ observable, probing $\mathcal{CP}$-symmetry breaking in strong interactions and topological properties of the QCD vacuum. Using a multiphase transport model with tunable CME strengths, we systematically compare Au+Au and isobar collisions at $\sqrt{s_{_{\rm NN}}} = 200$ GeV. We observe a reduced difference in the CME signal-to-background ratio between the spectator and participant planes for Au+Au collisions compared to isobar collisions. A comprehensive chi-square analysis across all three collision systems reveals stronger CME signatures in Au+Au collisions relative to isobar collisions, particularly when measured with respect to the spectator plane. Our findings demonstrate an enhanced experimental reliability of the two-plane method for the CME detection in Au+Au collisions.

\end{abstract}

%\pacs{25.75.-q, 25.75.Gz, 25.75.Nq}

  \maketitle

\section{Introduction}
\label{sec:intro}

Relativistic heavy-ion collisions create a unique environment where a quark-gluon plasma with strong collectivity is formed~\cite{Kolb:2000sd,Teaney:2000cw,Yan:2017ivm,Shen:2020mgh,Song:2017wtw,Lan:2022rrc,Wu:2021xgu,Shou:2024uga}, accompanied by the strongest known magnetic field generated by spectator protons from the colliding nuclei~\cite{Skokov:2009qp,Bzdak:2011yy,Deng:2012pc,Zhao:2017rpf,Zhao:2019ybo,Chen:2021nxs,Chen:2019qoe,Zhao:2022dac}. This environment provides an ideal laboratory for studying the topological properties of the QCD vacuum and anomalous chiral transport phenomena under extreme magnetic field conditions. Notably, the chiral magnetic effect (CME), which induces electric charge separation along the magnetic field direction in systems with chiral imbalance, serves as a crucial probe for detecting these phenomena~\cite{Kharzeev:2004ey,Kharzeev:2007jp,Fukushima:2008xe}.

The charge-dependent azimuthal correlation, $\gamma_{\alpha\beta}=\langle \cos(\phi_{\alpha}+\phi_{\beta}-2\Psi_{\rm RP})\rangle$, was initially proposed as a potential observable for detecting the CME~\cite{Voloshin:2004vk}. In this correlator, $\phi_{\alpha(\beta)}$ denotes the azimuthal angle of a charged particle $\alpha(\beta)$, and $\Psi_{\rm RP}$ represents the angle of the reaction plane. The difference between opposite-charge and same-charge correlations is represented by $\Delta\gamma$. Early measurements of this correlation by the STAR Collaboration~\cite{STAR:2009wot,STAR:2009tro,STAR:2013ksd,STAR:2014uiw} at RHIC and the ALICE Collaboration~\cite{ALICE:2012nhw} at the LHC aligned with CME expectations. However, significant background effects, especially those arising from elliptic flow, influenced the measured correlator~\cite{Bzdak:2010fd,Liao:2010nv,Schlichting:2010qia,Wang:2009kd,Wu:2022fwz,STAR:2013zgu}. Recent RHIC-STAR measurements have limited the CME fraction to below 10\%~\cite{Wang:2018ygc,Zhao:2020utk,Zhao:2019hta,Wang:2022eoo} in Au+Au collisions at $\sqrt{s_{_{\rm NN}}} = 200$ GeV. To separate the potential CME signal from the dominant background, several methods have been proposed~\cite{Zhao:2019hta,Wang:2022eoo,Li:2020dwr,Qi-Ye:2023zyf,Xu:2023elq,Zhao:2021yjo}. One promising approach involves using isobar collisions, which consist of systems with the same nucleon number but different proton numbers, such as $_{44}^{96}\textrm{Ru}+_{44}^{96}\textrm{Ru}$ and $_{40}^{96}\textrm{Zr}+_{40}^{96}\textrm{Zr}$ collisions~\cite{Voloshin:2010ut,Deng:2016knn}. The latest STAR evaluation of the CME signal established an upper limit of approximately 10\% for the CME fraction in the $\Delta\gamma$ measurement at a 95\% confidence level in isobar collisions at $\sqrt{s_{_{\rm NN}}} = 200$ GeV, after correcting for nonflow contamination~\cite{STAR:2023gzg,STAR:2023ioo,Chen:2024aom}. However, the nuclear structure effect poses a challenge for searching for the CME in isobar collisions, as it introduces differences in the backgrounds between the two isobar systems~\cite{Xu:2021vpn,Xu:2021uar,Zhang:2021kxj,Jia:2021oyt,Giacalone:2021udy,Jia:2021tzt,Jia:2021qyu,Zhao:2022grq,Li:2022bhl,Jia:2022ozr,Wang:2023yis}. In the newly developed chiral anomaly transport (CAT) module based on the state-of-the-art model based on a multiphase transport (AMPT) model，the upper limit of the CME signal is 15\% in isobar collisions~\cite{Yuan:2023skl,Yuan:2024wpz}, which aligns with the STAR data.

Numerous experimental observables have been employed to detect the true CME signal while minimizing background interference in Au+Au and isobar collisions. For this purpose, a two-plane measurement method, utilizing charge-dependent azimuthal correlations relative to the spectator plane (SP) and participant plane (PP), has been proposed~\cite{Xu:2017qfs,Xu:2017zcn}. This method is based on the fact that the background and CME signal exhibit different sensitivities or correlations to the two planes~\cite{Zhao:2019crj}. The STAR collaboration applied this method to quantify the fraction of the CME signal within the inclusive $\Delta\gamma$ correlation in both Au+Au and isobar collisions. For Au+Au collisions at $\sqrt{s_{_{\rm NN}}} = 200$ GeV, the STAR results suggest that the fraction of CME-induced charge separation is consistent with zero in peripheral centrality bins, while finite CME signals may exist in mid-central centrality bins~\cite{STAR:2021pwb}. This method is believed to eliminate most of the collective flow effect in the background; however, certain non-flow background effects require further investigation~\cite{Feng:2021pgf}. Furthermore, the spectator and participant plane methodology assumes that the ratio $a$ of elliptic flow relative to different reaction planes is equivalent to the ratio $b$ of CME signals relative to different reaction planes~\cite{STAR:2021pwb,Shi:2017cpu,Feng:2021pgf}. However, these two ratios might differ. In our previous study~\cite{Chen:2023jhx}, we calculated the CME signal-to-background ratio of $b$ over $a$ ($b/a = 0.65 \pm 0.18$) in isobar collisions at $\sqrt{s_{_{\rm NN}}} = 200$ GeV using the AMPT model with an initial CME signal, thus providing theoretical support for the experimental measurement of the CME in isobar collisions. In this work, the experimental search of CME signals in Au+Au collisions and the need to explore the values of $b$ across different collision systems motivated us to extend our calculations to Au+Au collisions at $\sqrt{s_{_{\rm NN}}} = 200$ GeV. We aim to simultaneously fit the experimental observables related to the CME in three different collision systems of Au+Au, Zr+Zr, and Ru+Ru within the AMPT model framework, thereby achieving synchronous constraint and extraction of the CME strengths in these three collision systems.

This paper is structured as follows. In Sec.~\ref{sec:model}, we describe the setup of the AMPT model with an initial CME signal and outline our two-plane method for extracting the fraction of the CME signal from the inclusive $\Delta\gamma$. Our model results are presented and compared with measurements from the STAR experiment in Sec.~\ref{sec:results}, where we also discuss the implications of our findings for the interpretation of experimental data and possible physical sources. Finally, Sec.~\ref{sec:summary} provides a summary of our main conclusions.   

\section{Model and method}
\label{sec:model}
\subsection{The AMPT model with initial CME signal}

The AMPT model is a multiphase transport framework designed to simulate the four main stages of relativistic heavy-ion collisions~\cite{Lin:2004en,Ma:2016fve,Lin:2021mdn}, which includes: 

(1) The HIJING model provides the initial condition. The transverse density profile of the colliding nucleus is modeled as a Woods-Saxon distribution. Multiple scatterings among participant nucleons generate the spatial and momentum distributions of minijet partons and soft excited strings. Using a string melting mechanism, the quark plasma is generated by melting parent hadrons. 

(2) Zhang's parton cascade (ZPC) model simulates the parton cascade stage. The ZPC model describes parton interactions via two-body elastic scatterings. The parton cross section is calculated using leading-order pQCD for gluon-gluon interactions. 

(3) A quark coalescence model combines two or three nearest quarks into hadrons to simulate hadronization. 

(4) A relativistic transport (ART) model simulates the stage of hadronic rescatterings, including resonance decays and all hadronic reactions involving elastic and inelastic scatterings among baryon-baryon, baryon-meson, and meson-meson interactions. 

Numerous previous studies have demonstrated that the AMPT model effectively describes various experimental observables in both large and small colliding systems at RHIC and the LHC~\cite{Lin:2004en,Ma:2016fve,Lin:2021mdn,Lin:2014tya,OrjuelaKoop:2015jss,Ma:2016bbw,He:2017tla,Huang:2021ihy,Chen:2022wkj,Chen:2022xpm}.   

Following the established methodology in Ref.~\cite{Ma:2011uma}, we implement a CME-like charge separation mechanism in the initial partonic stage of the AMPT model. By adjusting the percentage $p$, which defines the fraction of quarks participating in the CME-like charge separation, the CME signal strength can be controlled. The percentage $p$ is defined as follows:  

\begin{equation} \label{equ.01}   
p = \frac{N_{\uparrow(\downarrow)}^{+(-)}-N_{\downarrow(\uparrow)}^{+(-)}}{N_{\uparrow(\downarrow)}^{+(-)}+N_{\downarrow(\uparrow)}^{+(-)}},  
\end{equation} 
where $N$ represents the number of quarks of a given species (u, d, or s), $+$ and $-$ denote the positive and negative charges of quarks, and $\uparrow$ and $\downarrow$ indicate the quarks' moving direction which is parallel or anti-parallel to the magnetic field.

\subsection{Spectator and participant planes}
\label{sec:plane}
The two-plane method utilizes the distinct plane correlations, i.e., the elliptic flow-driven background is predominantly correlated with the participant plane, whereas the CME signal shows stronger correlation with the spectator plane~\cite{Xu:2017qfs,Xu:2017zcn}. The spectator and participant planes used therein can be reconstructed using the following equations:  

\begin{equation}\label{equ.02}  
	\psi_{\rm S P}=\frac{\operatorname{atan} 2\left(\left\langle r_{\rm n}^2 \sin \left(2 \phi_{\text {n }}\right)\right\rangle,\left\langle r_{\rm n}^2 \cos \left(2 \phi_{\text {n }}\right)\right\rangle\right)}{2},  
\end{equation}   
\begin{equation}\label{equ.03}  
	\psi_{\rm P P}=\frac{\operatorname{atan} 2\left(\left\langle r_{\rm p}^2 \sin \left(2 \phi_{\text {p }}\right)\right\rangle,\left\langle r_{\rm p}^2 \cos \left(2 \phi_{\text {p }}\right)\right\rangle\right)+\pi}{2},  
\end{equation} 
where $r_{\rm n}$ and $\phi_{\rm n}$ represent the displacement and azimuthal angle of spectator neutrons in the transverse plane, respectively, while $r_{\rm p}$ and $\phi_{\rm p}$ represent the displacement and azimuthal angle of participating partons in the transverse plane. All spatial information regarding the displacement and azimuthal angle is obtained from the initial state of the AMPT model. The $\langle ... \rangle$ means averaged over all partons for each event. The participant plane (PP) can be experimentally determined through the event plane reconstructed from final-state hadrons. In this study, the PP is employed to minimize the nonflow effect ~\cite{Feng:2021pgf} in the reconstructed $\psi_{\mathrm{EP}}$. However, we have systematically verified that all conclusions remain quantitatively consistent when employing the event plane (EP) method. The corresponding elliptic flow coefficients, $v_2\{\mathrm{SP}\}$ and $v_2\{\mathrm{PP}\}$, using the spectator plane ($\psi_{\mathrm{SP}}$) and participant plane ($\psi_{\mathrm{PP}}$) methods, respectively, through the following definitions:   

\begin{equation}\label{equ.04}  
	v_2\{\rm SP\}=\left\langle\left\langle\cos 2\left(\phi-\psi_{\mathrm{SP}}\right)\right\rangle\right\rangle,  
\end{equation} 
\begin{equation}\label{equ.05}  
	v_2\{\rm PP\}=\left\langle\left\langle\cos 2\left(\phi-\psi_{\mathrm{PP}}\right)\right\rangle\right\rangle.  
\end{equation} 
where $\phi$ represent the azimuthal angle of final hadrons in the transverse momentum plane and $\langle\langle ... \rangle\rangle$ means averaged over all charged hadrons for all events.
\begin{figure}
	\includegraphics[scale=0.4]{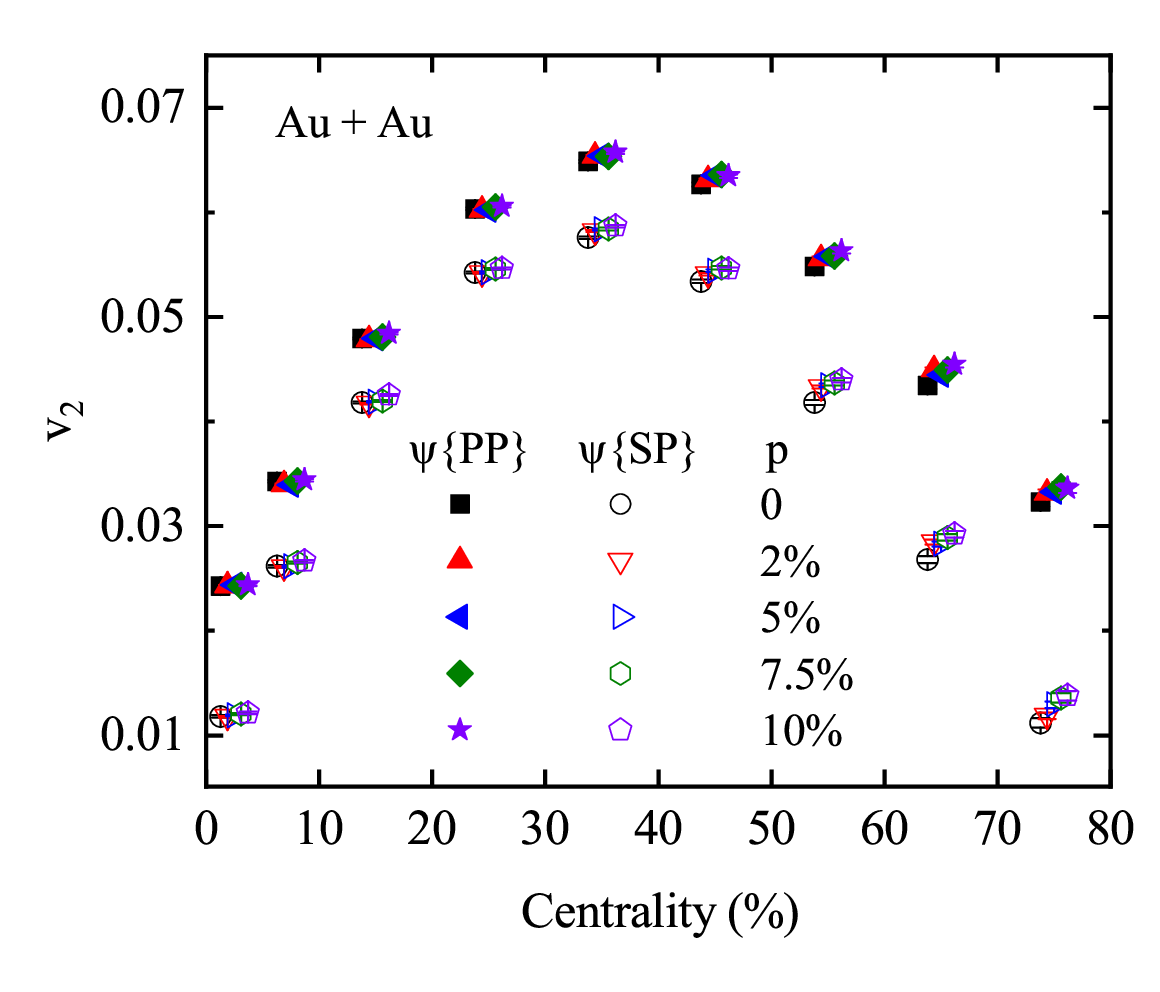}
	\caption{(Color online)  AMPT results on centrality dependence of elliptic flow $v_2\{\mathrm{PP}\}$ (solid symbols) and $v_2\{\mathrm{SP}\}$ (open symbols) in Au+Au collisions at $\sqrt{s_{_{\rm NN}}} = 200$ GeV from the AMPT model with different strengths of the CME. The data points are shifted along the $x$ axis for clarity.}
	\label{fig-01}
\end{figure}

Figure~\ref{fig-01} presents the centrality dependence of $v_2\{\mathrm{PP}\}$ and $v_2\{\mathrm{SP}\}$ for charged hadrons with $0.2<p_T<2.0 ~\mathrm{GeV} / c$ and $|\eta|<1$, obtained from the AMPT model with varying CME strengths in Au+Au collisions at $\sqrt{s_{_{\rm NN}}} = 200$ GeV. As anticipated, $v_2\{\mathrm{PP}\}$ is consistently larger than $v_2\{\mathrm{SP}\}$ in all cases, as elliptic flow is more strongly correlated with the participant plane than with the spectator plane. The values of both $v_2\{\mathrm{PP}\}$ and $v_2\{\mathrm{SP}\}$ are largely unaffected by the increasing strength of the CME signal.

\begin{figure}
	\includegraphics[scale=0.4]{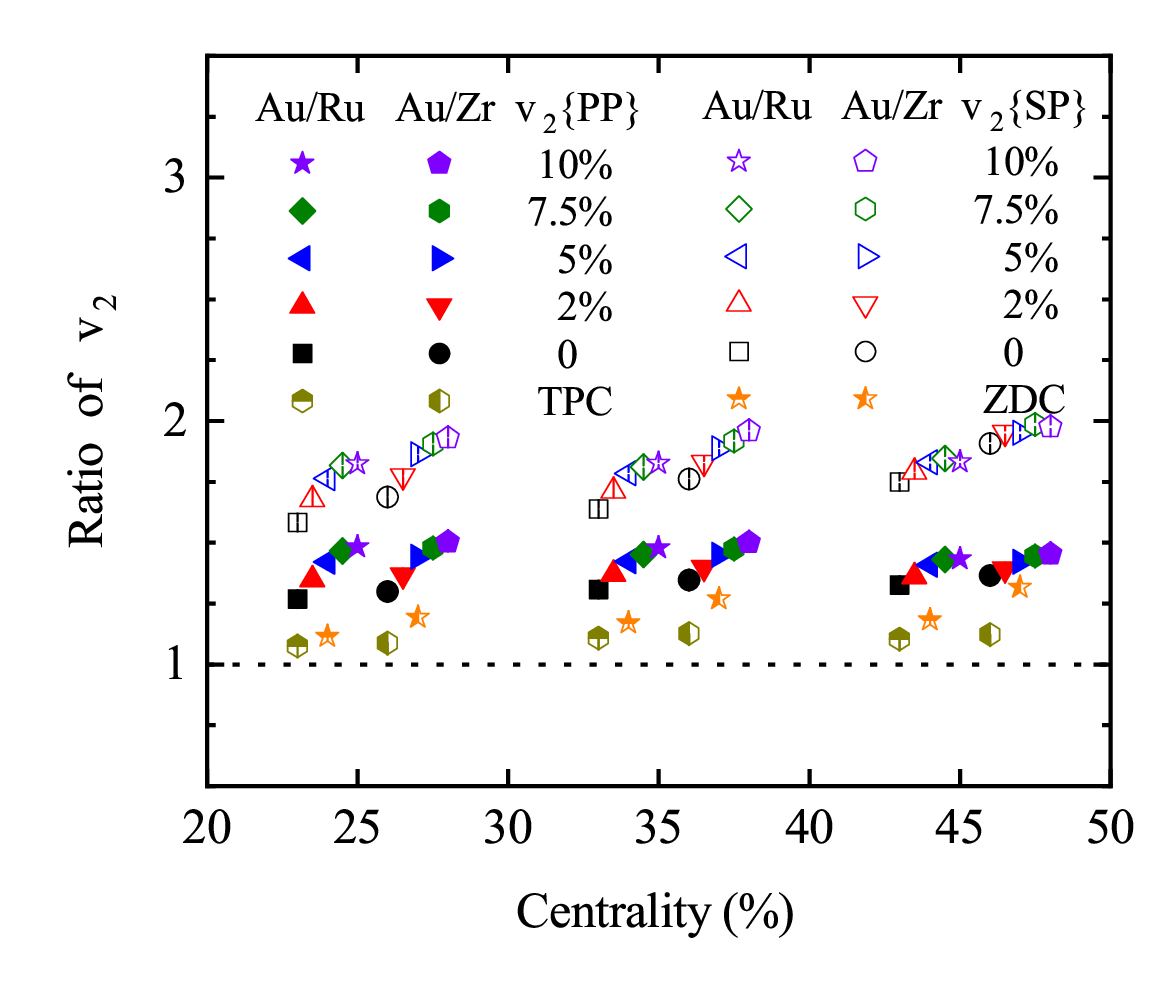}
	\caption{(Color online) The centrality dependence of elliptic flow $v_2\{\mathrm{PP}\}$ (solid symbols) and $v_2\{\mathrm{SP}\}$ (open symbols) ratios of Au+Au collisions to Ru+Ru collisions and Zr+Zr collisions, respectively, at $\sqrt{s_{_{\rm NN}}} = 200$ GeV from the AMPT model with different strengths of the CME, in comparison with the STAR data~\cite{STAR:2021pwb,STAR:2021mii}. The data points are shifted along the $x$ axis for clarity.}
	\label{fig-02}
\end{figure}

Figure~\ref{fig-02} presents the ratios of $v_2\{\mathrm{PP}\}$ and $v_2\{\mathrm{SP}\}$ for Au+Au collisions relative to Ru+Ru and Zr+Zr collisions. Note that all the calculations about Ru+Ru and Zr+Zr collisions are taken from Ref.~\cite{Chen:2023jhx}. Since the CME is more likely to occur at centrality bins of 20-50\% and to avoid large errors, the comparison is restricted to these centrality bins. The ratio of $v_2\{\mathrm{PP}\}$ is observed to be larger than that of $v_2\{\mathrm{SP}\}$, consistent with experimental trends. Compared to the experimental data, the relatively larger ratio in our results arises because the $v_2\{\mathrm{PP}\}$ and $v_2\{\mathrm{SP}\}$ values calculated from our isobar collision simulations are smaller than the experimental results. Similarly, the $v_2\{\mathrm{PP}\}$ and $v_2\{\mathrm{SP}\}$ ratios of Au+Au collisions to Ru+Ru collisions are smaller than those of Au+Au collisions to Zr+Zr collisions, reflecting the influence of nuclear structure. This indicates that the $v_2\{\mathrm{PP}\}$ and $v_2\{\mathrm{SP}\}$ values for Ru+Ru collisions are larger than those for Zr+Zr collisions due to the nuclear structure effect~\cite{Zhao:2022grq,Chen:2023jhx}.

\subsection{Two-plane method to extract $f_{\rm CME}$}
\label{sec:fcme}
This subsection presents the original two-plane method for detecting and extracting of the fraction of the CME signal and its optimization of the method through the AMPT model. The experimentally measurable CME observable, denoted as $\Delta\gamma$, consists of the CME signal and the background effect. These background effects are predominantly attributed to elliptic flow and non-flow effects, which originate from resonance decays, jet correlations, and so on. Consequently, the experimentally measured observable with respect to different planes can be mathematically expressed as the sum of two components:

\begin{equation}\label{equ.05}  
	\Delta \gamma \{\mathrm{\psi}\}=\Delta \gamma_{\rm Bkg}\{\mathrm{\psi}\}+\Delta \gamma_{\rm CME}\{\mathrm{\psi}\}  
\end{equation}  
where $\psi$ represents either the spectator plane ($\psi_{\mathrm{SP}}$) or participant plane ($\psi_{\mathrm{PP}}$). The ratios of the elliptic flow and the measured observable with respect to two different planes are defined as $a$ and $A$, respectively, as follows:  

\begin{equation}\label{equ.07}  
	a=v_{2}\{\mathrm{SP}\} / v_{2}\{\mathrm{PP}\},  
\end{equation} 

\begin{equation}\label{equ.08}  
	A=\Delta \gamma{\{\mathrm{SP}\}} / \Delta \gamma{\{\mathrm{PP}\}}.  
\end{equation} 

The parameter $a$ is expected to be governed by the two-plane correlation factor, which can be quantitatively expressed through the following relationship, i.e., $a = \left\langle\cos 2\left(\psi_{\mathrm{PP}}-\psi_{\mathrm{SP}}\right)\right\rangle$~\cite{Xu:2017qfs,Bloczynski:2012en,Shi:2019wzi}. Through a straightforward mathematical transformation, the fraction contribution of the CME signal to the total measured observable, denoted as $f_{\rm CME}$, can be expressed as: 

\begin{equation}\label{equ.09}  
	f_{\rm CME}=\frac{\Delta \gamma_{\rm CME}\{\mathrm{PP}\}}{\Delta \gamma\{\mathrm{PP}\}}=\frac{A / a-1}{1 / a^2-1}.  
\end{equation}   

Equation~(\ref{equ.09}) shows that the fraction of the CME signal within the measured CME observable can be determined by measuring $A$ and $a$. However, as critically noted in Refs.~\cite{Shi:2017cpu,Feng:2021pgf}, a potential discrepancy may exist between the ratio of the CME signal and the inverse ratio of the elliptic flow. This discrepancy has been quantitatively verified through our recent AMPT calculation for isobar collisions~\cite{Chen:2023jhx}. Therefore, in the more general case, the following relationship should hold:
\begin{equation}\label{equ.10}  
	\Delta \gamma_{\mathrm{CME}}\{\mathrm{PP}\}=b \Delta \gamma_{\mathrm{CME}}\{\mathrm{SP}\}.  
\end{equation} 
where $b$ represents the ratio of the CME signals with respect to different planes. In this way, the following equation can be derived:   
\begin{equation}\label{equ.11}  
	\Delta \gamma \{\mathrm{SP}\} = a \Delta \gamma_{\rm Bkg}\{\mathrm{PP}\} + \frac{\Delta \gamma_{\rm CME}\{\mathrm{PP}\}}{b}.  
\end{equation}
After accounting for $b$, a more realistic estimation of the fraction of the CME signal to the total observable can be expressed through the following modified relation:  

\begin{equation}\label{equ.12}  
	f_{\rm CME}\{b\}=\frac{\Delta \gamma_{\rm CME}\{\mathrm{PP}\}}{\Delta \gamma\{\mathrm{PP}\}}=\frac{A / a-1}{1 / a b-1}.
\end{equation}   
The remaining task is to calculate $b$, which can be determined theoretically. In the AMPT model, the CME signal is simulated by introducing a certain percentage $p$ of partons in the initial state to participate in the charge separation phenomenon. Therefore, the value of $b$ can be determined in the AMPT model using the following relation:  

\begin{equation}\label{equ.13}  
b =\frac{\Delta \gamma\{\mathrm{PP}\}(p \neq 0) - \Delta \gamma\{\mathrm{PP}\}(p=0)}{\Delta \gamma\{\mathrm{SP}\}(p \neq 0) - \Delta \gamma\{\mathrm{SP}\}(p=0)},  
\end{equation}  
where the numerator and denominator represent the CME signal inside the measured CME observable with respect to the participant and spectator planes, respectively.

\section{Results and Discussions}
\label{sec:results}

This section presents the AMPT model results, focusing on charge-dependent azimuthal correlations for charged particles relative to both the spectator and participant planes. To compare with the measurements from the STAR experiment, we use kinetic cuts of $0.2<p_T<2.0$ GeV/$c$ and $|\eta|<1$, consistent with the STAR experimental setup.  

\begin{figure}
	\includegraphics[scale=0.4]{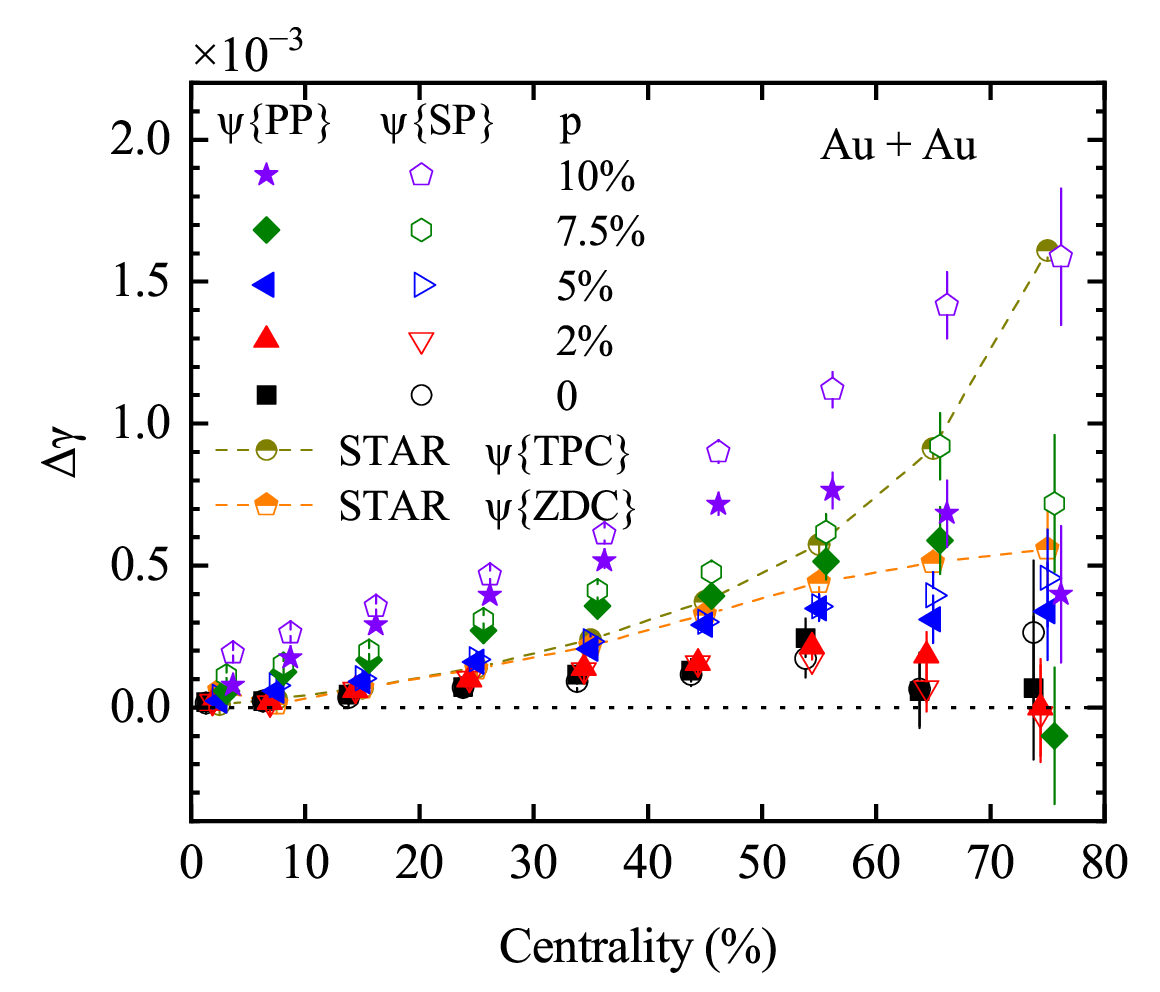}
	\caption{(Color online) The centrality dependence of $\Delta \gamma\{\mathrm{PP}\}$ (solid symbols) and $\Delta \gamma\{\mathrm{SP}\}$ (open symbols) in Au+Au collisions at $\sqrt{s_{_{\rm NN}}} = 200$ GeV from the AMPT model with different strengths of the CME, in comparison with the STAR data linked with dotted line~\cite{STAR:2021pwb}. The data points are shifted along the $x$ axis for clarity.}
	\label{fig-03}
\end{figure}

Figure~\ref{fig-03} presents the centrality dependence of $\Delta \gamma\{\mathrm{PP}\}$ and $\Delta \gamma\{\mathrm{SP}\}$ from the AMPT model with varying CME strengths in Au+Au collisions at $\sqrt{s_{_{\rm NN}}} = 200$ GeV. Compared to the STAR data, the CME signal percentage at $p = 5\%$ or $7.5\%$ aligns more closely with the experimental results. Notably, $\Delta \gamma\{\mathrm{SP}\}$ exceeds $\Delta \gamma\{\mathrm{PP}\}$, indicating that the spectator plane serves as a more sensitive probe of the CME because of its stronger correlation with the magnetic field direction compared to with respect to the participant plane. 

\begin{figure}
	\includegraphics[scale=0.4]{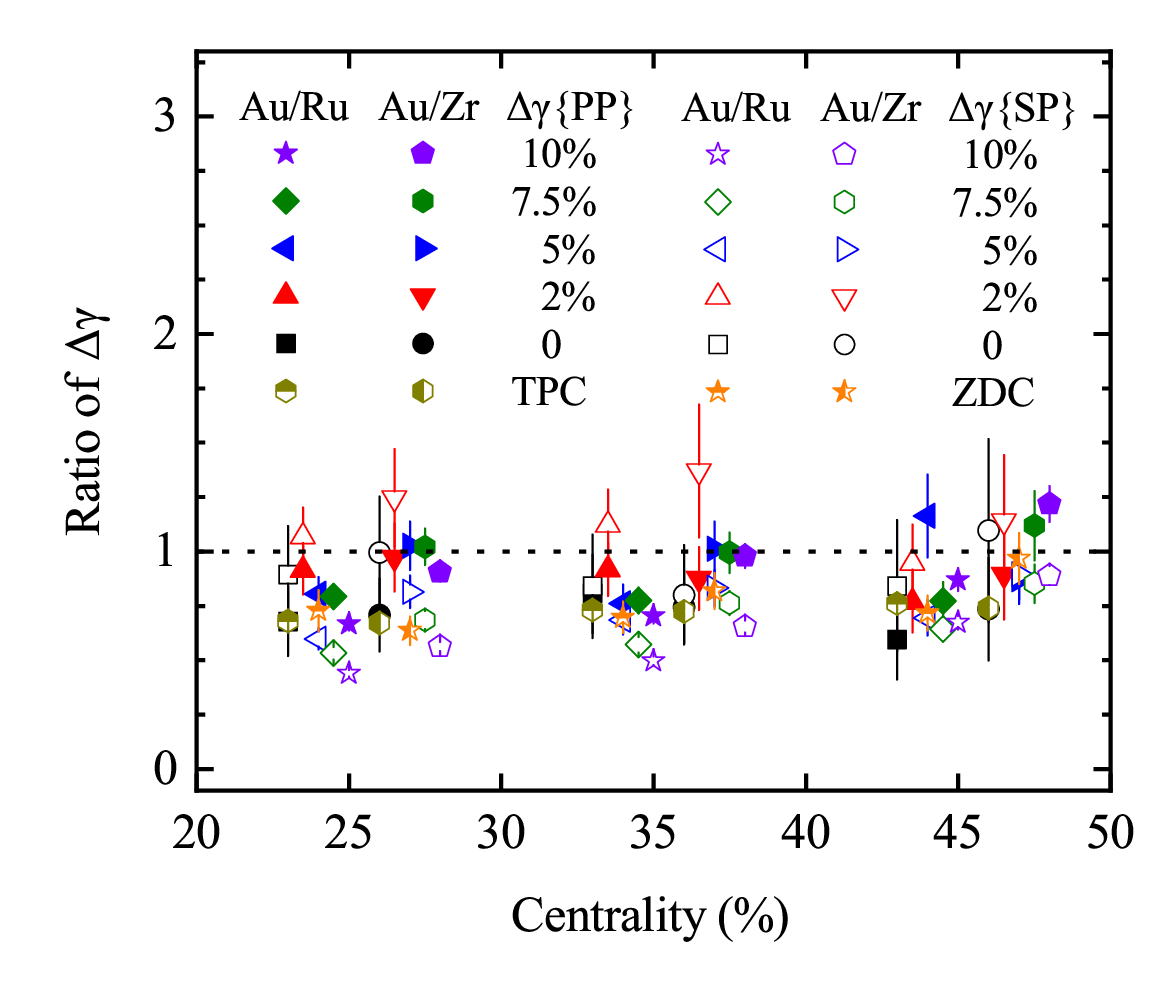}
	\caption{(Color online) The centrality dependence of $\Delta \gamma\{\mathrm{PP}\}$ (solid symbols)  and $\Delta \gamma\{\mathrm{SP}\}$ (open symbols) ratios of Au+Au collisions to Ru+Ru collisions and Zr+Zr collisions, respectively, at $\sqrt{s_{_{\rm NN}}} = 200$ GeV from the AMPT model with different strengths of the CME, in comparison with the STAR data~\cite{STAR:2021pwb,STAR:2021mii}. The data points are shifted along the $x$ axis for clarity.} 
	\label{fig-04}
\end{figure}

Figure~\ref{fig-04} presents the centrality dependence of the $\Delta \gamma\{\mathrm{PP}\}$ and $\Delta \gamma\{\mathrm{SP}\}$ ratios for Au+Au collisions relative to Ru+Ru and Zr+Zr collisions. Considering the effect of errors, both experimental and theoretical results indicate that the $\Delta \gamma$ value for Au+Au collisions is smaller than that for isobar collisions, as most of the ratios are less than one. For the ratio of $\Delta \gamma\{\mathrm{PP}\}$, the AMPT results without the CME signal agree more closely with the experimental results, whereas for the ratio of $\Delta \gamma\{\mathrm{SP}\}$, the AMPT results with the CME signal show better agreement with the experimental results. Since the CME signal is correlated more strongly with the SP than the PP, we observe that the ratio of $\Delta \gamma\{\mathrm{SP}\}$ decreases with increasing CME strength. This suggests that $\Delta \gamma\{\mathrm{SP}\}$ in isobar collisions grows more rapidly than that in Au+Au collisions as the CME strength becomes stronger.

\begin{figure}
	\includegraphics[scale=0.40]{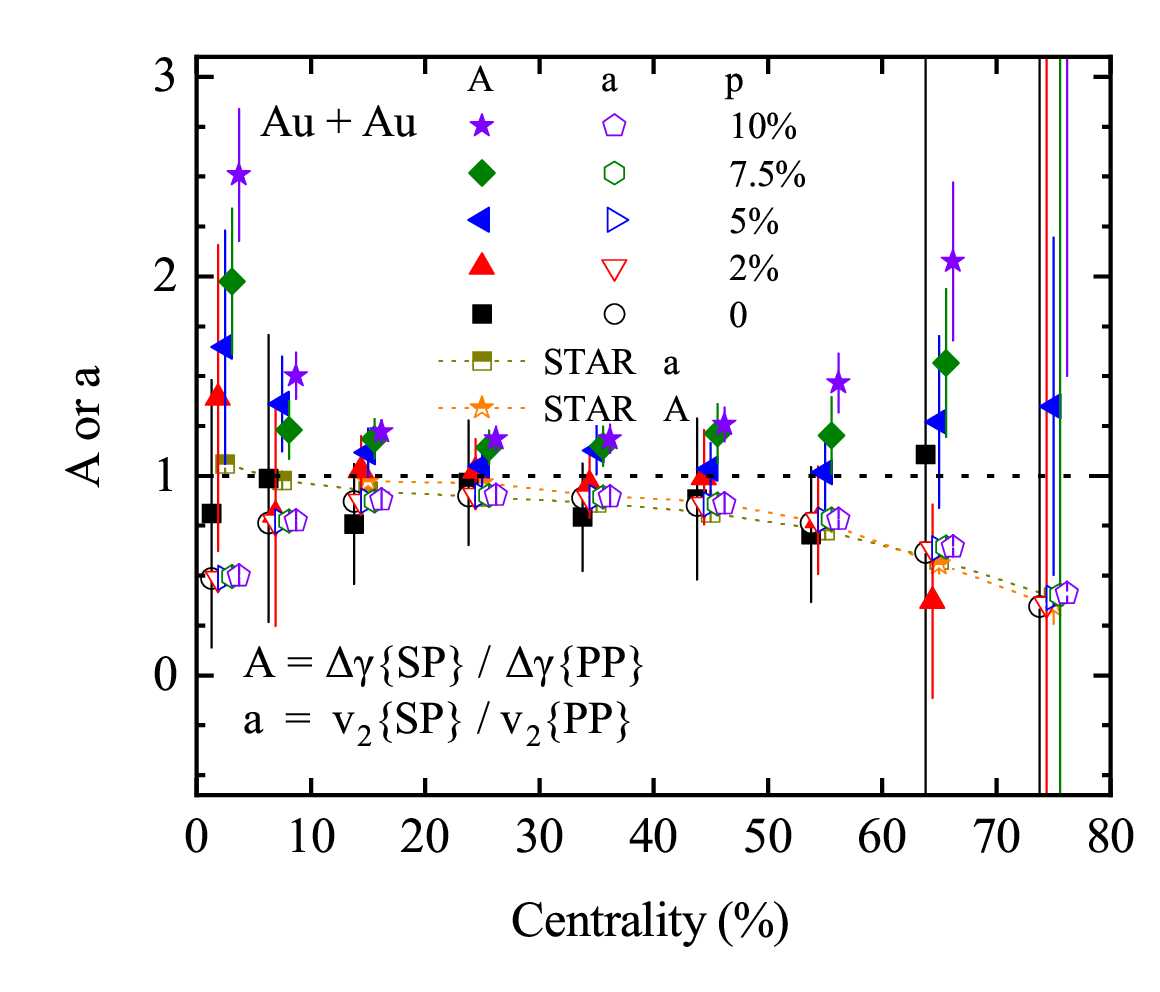}
	\caption{(Color online) The centrality dependence of $A$ (solid symbols) and $a$ (open symbols) in Au+Au collisions at $\sqrt{s_{_{\rm NN}}} = 200$ GeV from the AMPT model with different strengths of the CME, in comparison with the STAR data linked with dotted line~\cite{STAR:2021pwb}. The data points are shifted along the $x$ axis for clarity.}
	\label{fig-05}
\end{figure}

Figure~\ref{fig-05} shows the centrality dependence of $A=\Delta \gamma{\{\mathrm{SP}\}} / \Delta \gamma{\{\mathrm{PP}\}}$ and $a=v_{2}\{\mathrm{SP}\} / v_{2}\{\mathrm{PP}\}$ from the AMPT model with varying CME strengths, compared to STAR experimental data~\cite{STAR:2021pwb}. As the CME strength in the AMPT model increases, the value of $a$ remains nearly constant and consistently below unity. It is verified that $a$ follows the expectation $a = \left\langle\cos 2\left(\psi_\mathrm{PP}-\psi_\mathrm{SP}\right)\right\rangle$, indicating that the CME impacts $v_2$ similarly for the two different planes. In contrast, the values of $A$ increase with $p$, suggesting that the CME affects $\Delta\gamma$ differently for the two different planes. For small values of $p$, the model results align more closely with the experimental data.

\begin{figure}
	\includegraphics[scale=0.40]{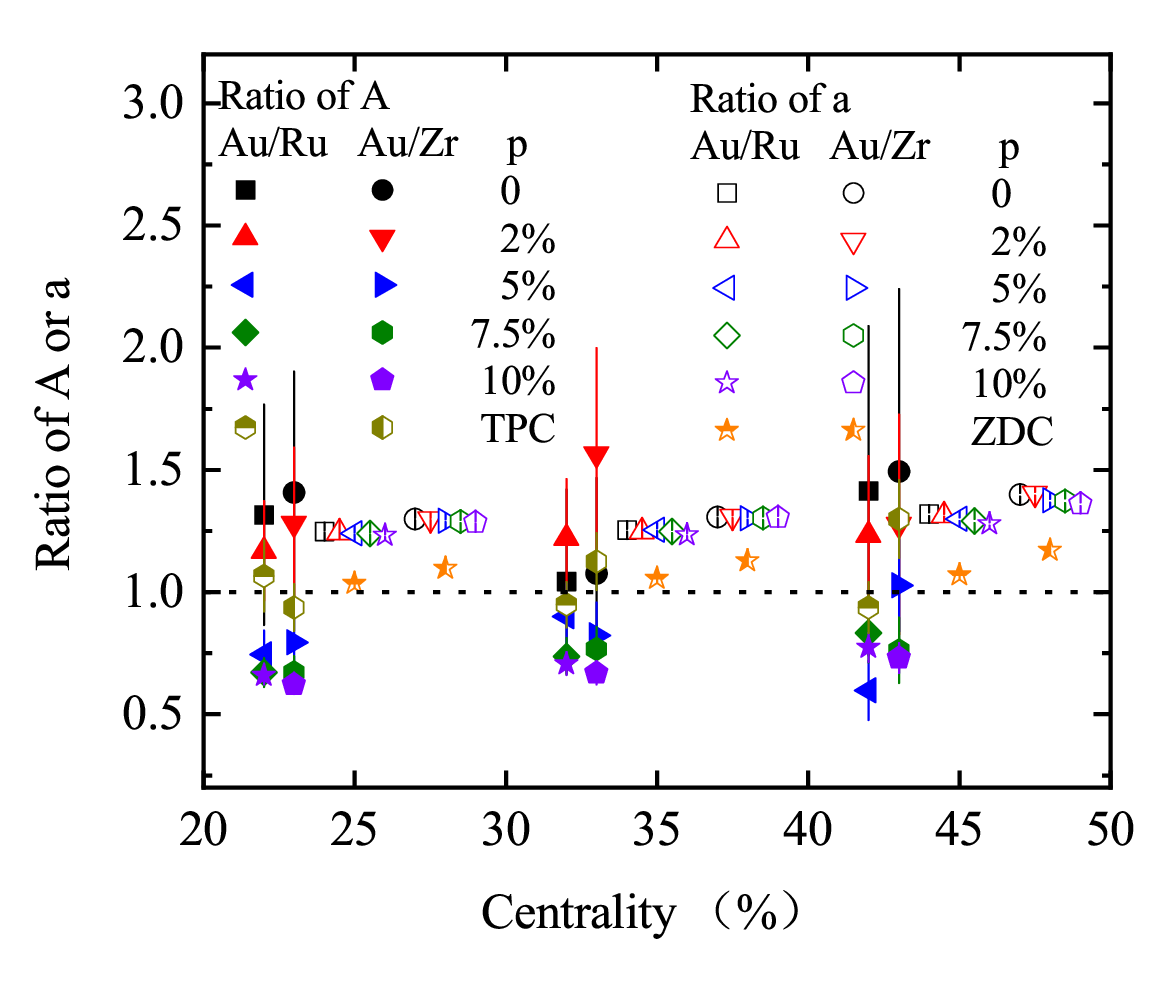}
	\caption{(Color online) The centrality dependence of $A$ (solid symbols) and $a$ (open symbols) ratios of Au+Au collisions to Ru+Ru collisions and Zr+Zr collisions，respectively, at $\sqrt{s_{_{\rm NN}}} = 200$ GeV from the AMPT model with different strengths of the CME, in comparison with the STAR data~\cite{STAR:2021pwb,STAR:2021mii}. The data points are shifted along the $x$ axis for clarity.}
	\label{fig-06}
\end{figure}

Figure~\ref{fig-06} shows the centrality dependence of the $A$ and $a$ ratios for Au+Au collisions relative to Ru+Ru and Zr+Zr collisions. The ratio of $a$ remains nearly unchanged, consistent with the experimental data within errors. The $a$ ratio greater than one indicates that the $ \left\langle\cos 2\left(\psi_\mathrm{PP}-\psi_\mathrm{SP}\right)\right\rangle$ value is larger, implying a smaller difference angle between $\psi_\mathrm{PP}$ and $\psi_\mathrm{SP}$ for Au+Au collisions compared to isobar collisions. For larger CME strengths, the ratio of $A$ is less than one, indicating that the ratios of $\Delta \gamma\{\mathrm{SP}\}$ and $\Delta \gamma\{\mathrm{PP}\}$ are smaller in Au+Au collisions than in isobar collisions, i.e., $\Delta \gamma\{\mathrm{SP}\}$ and $\Delta \gamma\{\mathrm{PP}\}$ are closer in Au+Au collisions than those in isobar collisions.   

\begin{figure}
	\includegraphics[scale=0.40]{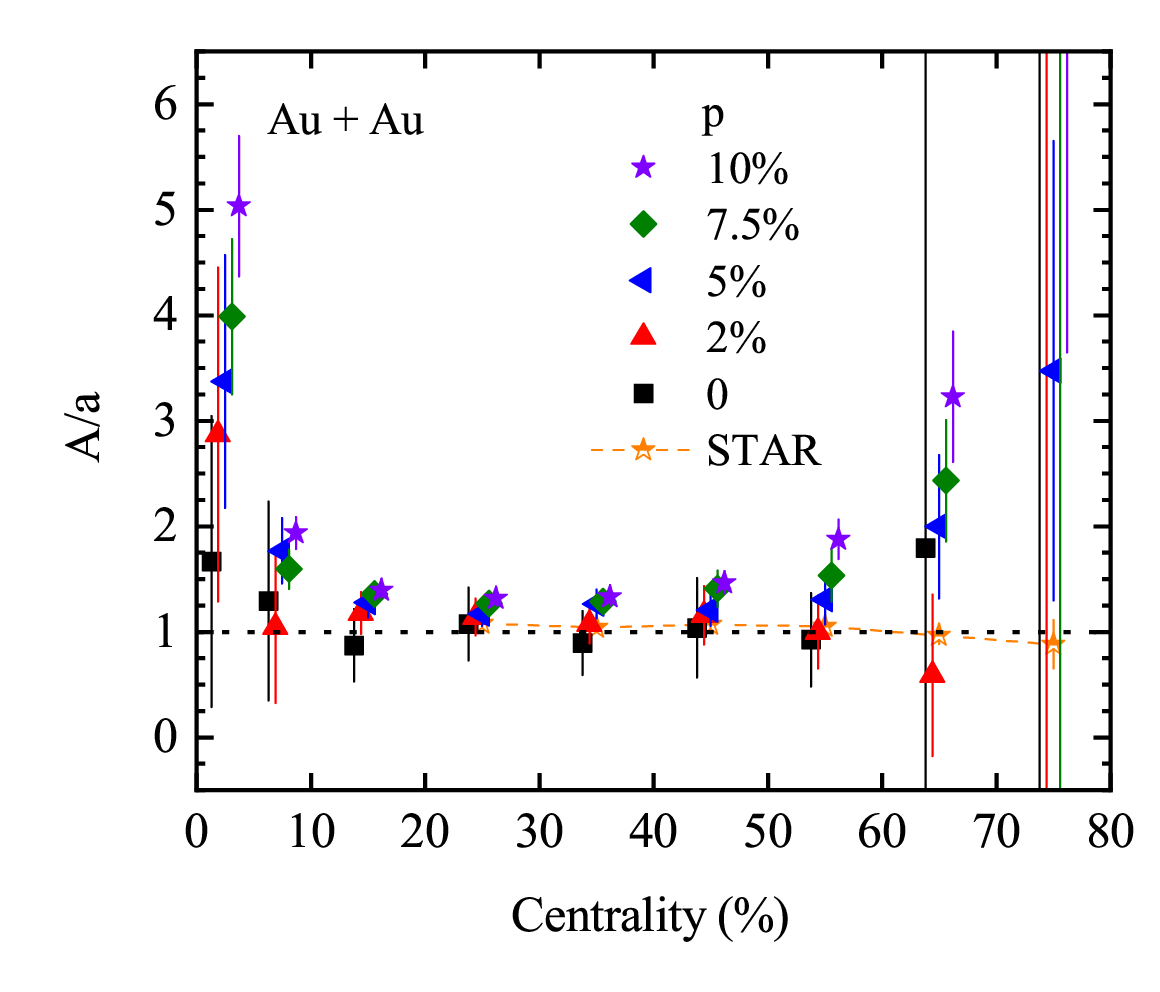}
	\caption{(Color online) The centrality dependence of $A/a$ in Au+Au collisions at $\sqrt{s_{_{\rm NN}}} = 200$ GeV from the AMPT model with different strengths of the CME, in comparison with the STAR data linked with dotted line~\cite{STAR:2021pwb}. The data points are shifted along the $x$ axis for clarity.}
	\label{fig-07}
\end{figure}

Figure~\ref{fig-07} presents $A/a$ as a function of centrality, predicted by the AMPT model with varying CME strengths. According to Eqs.~(\ref{equ.09}) and (\ref{equ.12}), $A/a$ values greater than unity indicate the presence of a CME signal within the $\Delta\gamma$ observable. Focusing on mid-central collisions (20-50\% centrality), where the CME effect is more measurable, $A/a>1$ is observed for all cases except when $p=0$ and $p=2\%$. Notably, $A/a$ increases with CME strength, suggesting that this ratio reflects the strength of the CME signal. The experimental data are closer to the cases with the lower strengths of the CME signal.

\begin{figure}
	\includegraphics[scale=0.40]{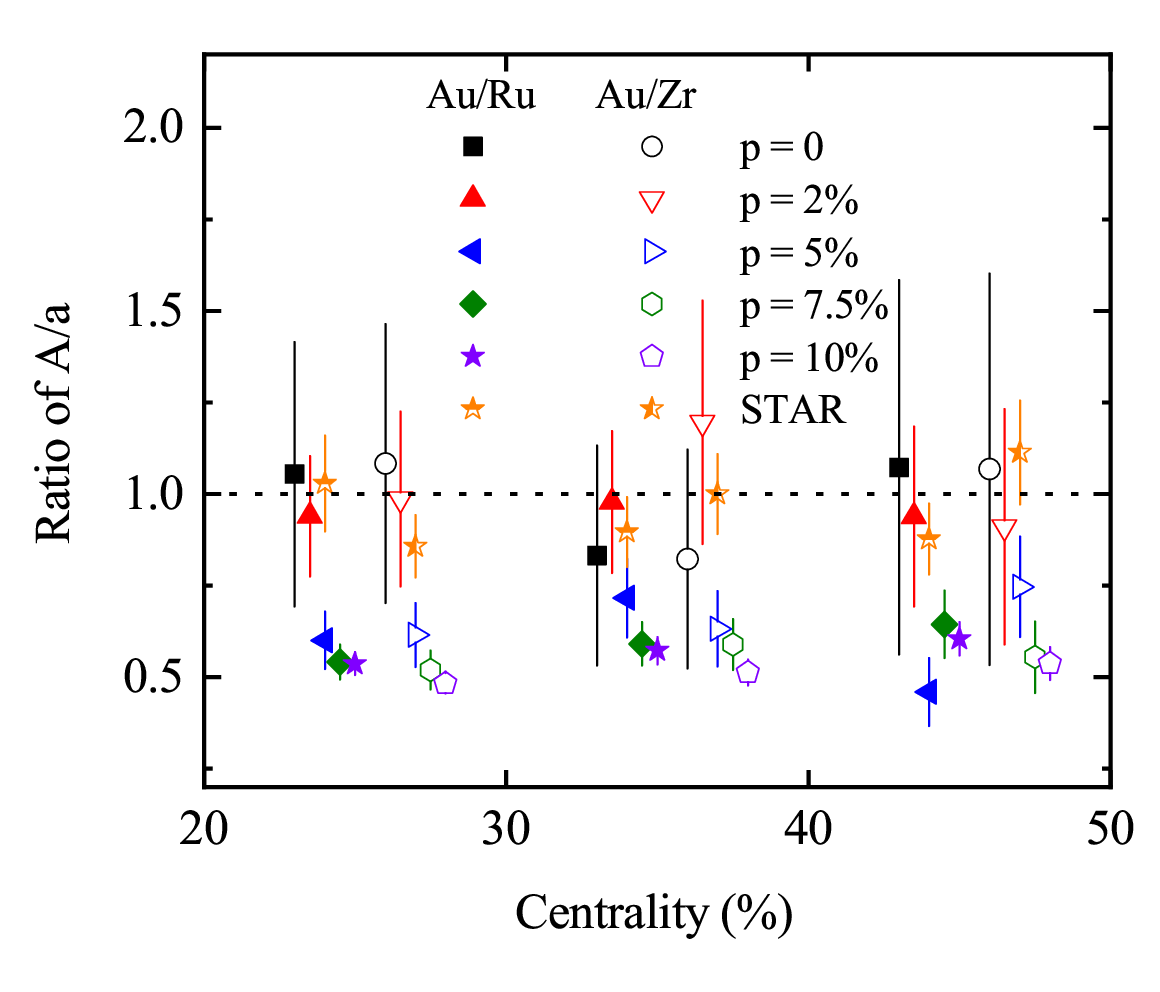}
	\caption{(Color online) The centrality dependence of $A/a$ ratios of Au+Au collisions to Ru+Ru collisions and Zr+Zr collisions, respectively, at $\sqrt{s_{_{\rm NN}}} = 200$ GeV from the AMPT model with different strengths of the CME, in comparison with the STAR data~\cite{STAR:2021pwb,STAR:2021mii}. The data points are shifted along the $x$ axis for clarity. }
	\label{fig-08}
\end{figure}

Figure~\ref{fig-08} shows the centrality dependence of the $A/a$ ratios for Au+Au collisions relative to Ru+Ru and Zr+Zr collisions. Focusing on mid-central collisions (20-50\% centrality), the $A/a$ ratios for Au+Au and isobar collisions exhibit a trend similar to the $A$ ratio shown in Fig.~\ref{fig-06}, driven by a smaller variation in $a$ and a larger variation in $A$. The experimental data are also closer to the cases with the lower strengths of the CME signal.

\begin{figure}
	\includegraphics[scale=0.40]{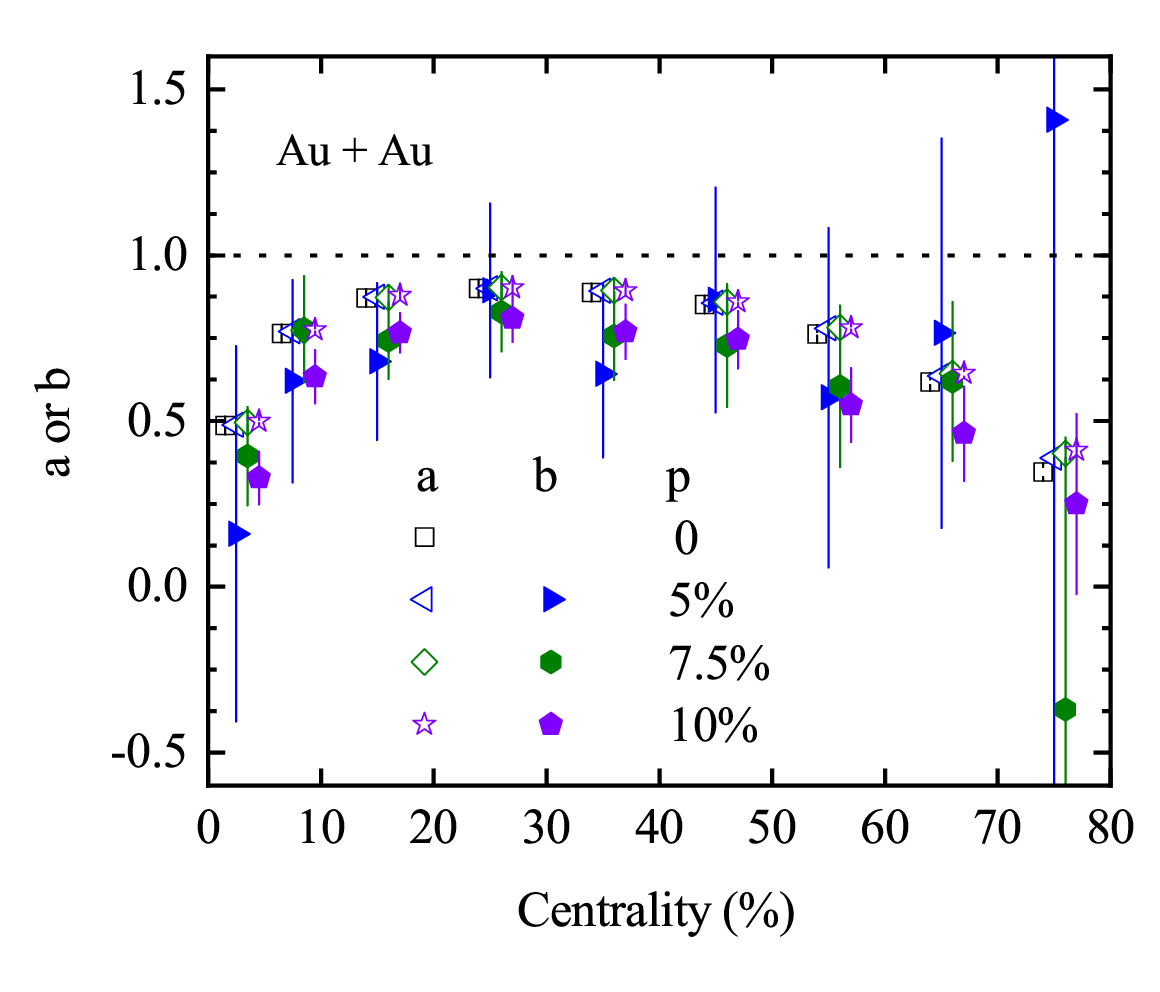}
	\caption{(Color online) AMPT results on centrality dependence of $a$ (solid symbols) and $b$ (open symbols) in Au+Au collisions at $\sqrt{s_{_{\rm NN}}} = 200$ GeV from the AMPT model with different strengths of the CME. The data points are shifted along the $x$ axis for clarity.}
	\label{fig-09}
\end{figure}

Figure~\ref{fig-09} shows the centrality dependence of $a$ and $b$, calculated using Eq.~(\ref{equ.13}), based on the AMPT model with varying CME strengths. Notably, $a$ and $b$ exhibit significant differences. In the 20-50\% centrality bins, $b$ remains consistently smaller than $a$. Moreover, $a$ shows minimal dependence on CME strength, consistent with the observations in Fig.~\ref{fig-05}. In contrast, $b$ reflects the capability of the PP method to capture the CME signal observed in the SP method. Although the statistical errors are substantial, $b$ shows no significant variation with CME strengths, except for the 2\% CME strength case, which is excluded due to its large statistical uncertainties.  

\begin{figure}
	\includegraphics[scale=0.40]{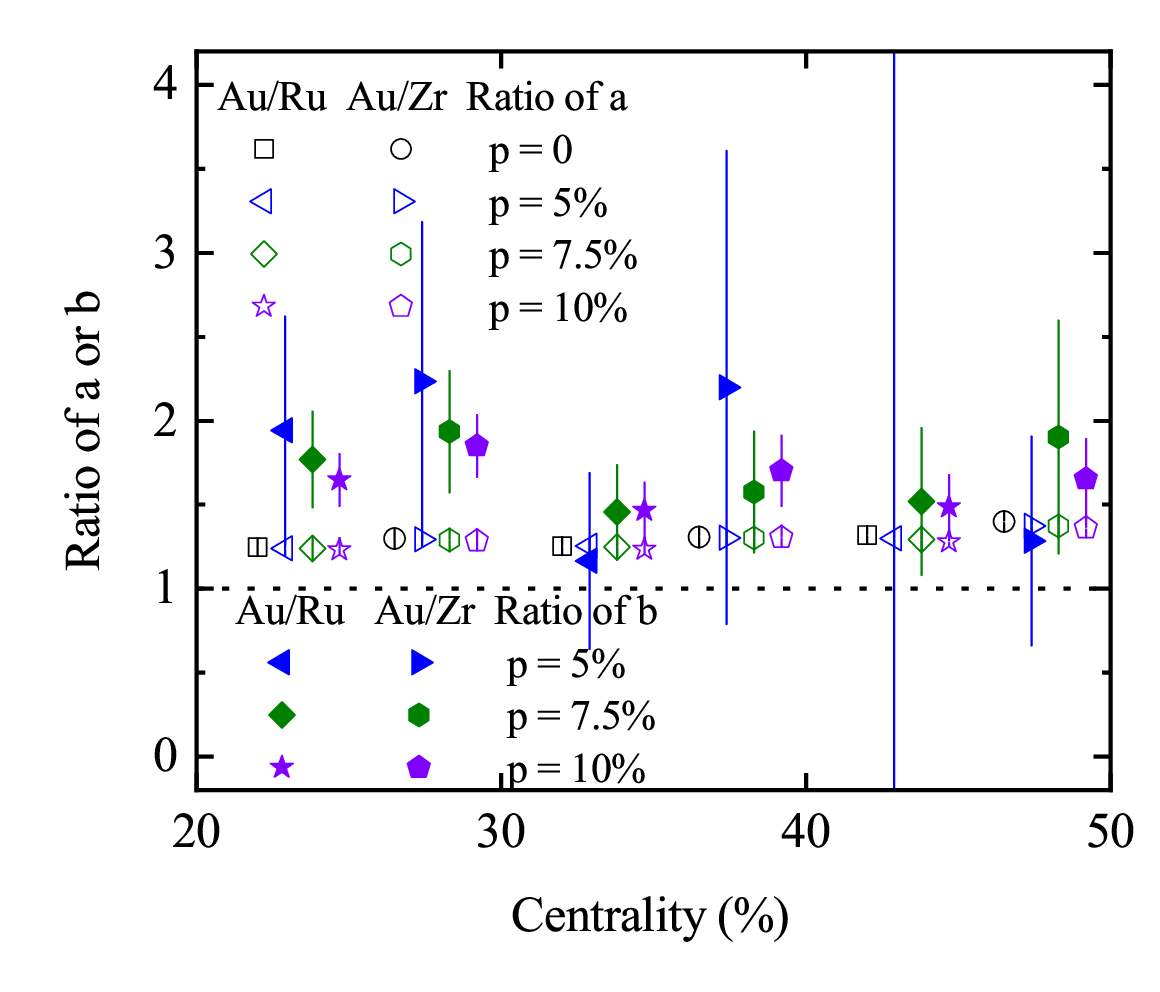}
	\caption{(Color online) AMPT results on centrality dependence of the  $a$ (open symbols) and $b$ (solid symbols) ratios of  Au+Au collisions to Ru+Ru collisions and Zr+Zr collisions, respectively, at $\sqrt{s_{_{\rm NN}}} = 200$ GeV from the AMPT model with different strengths of the CME. The data points are shifted along the $x$ axis for clarity.}
	\label{fig-10}
\end{figure}

Figure~\ref{fig-10} shows the centrality dependence of the $a$ and $b$ ratios for Au+Au collisions relative to Ru+Ru and Zr+Zr collisions, respectively, based on the AMPT model with varying CME strengths. Focusing on the 20-50\% centrality bins, the $a$ ratio remains nearly unchanged and is smaller than the $b$ ratio. The $b$ ratio decreases as the CME strength increases within errors. According to Eq.~(\ref{equ.13}), a larger $b$ value indicates a smaller difference in the net CME strength between the two planes. Therefore, our results indicate that the difference is smaller in Au+Au collisions than in isobar collisions.
 
\begin{figure}
	\includegraphics[scale=0.40]{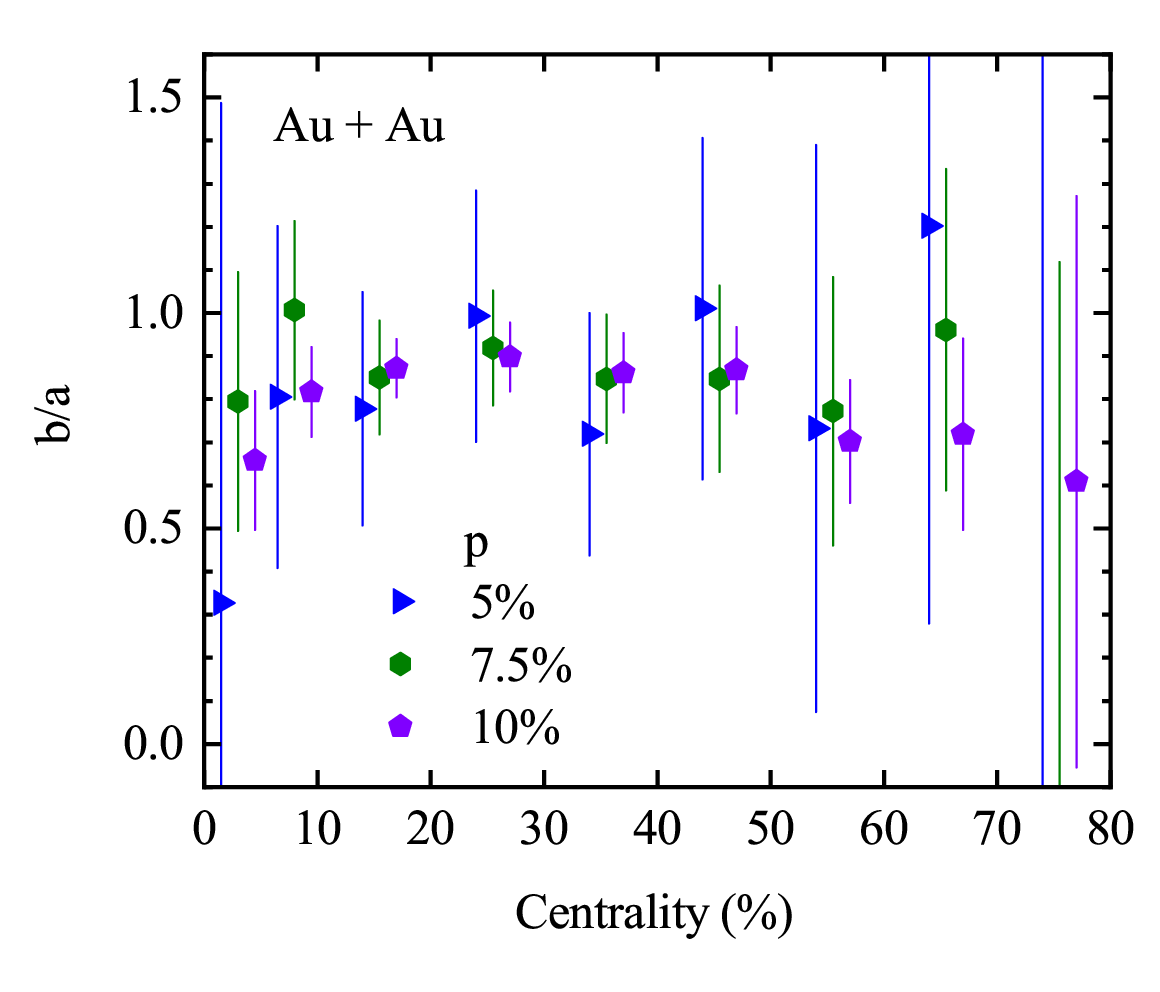}
	\caption{(Color online) The centrality dependence of $b/a$ in Au+Au collisions at $\sqrt{s_{_{\rm NN}}} = 200$ GeV from the AMPT model with different strengths of the CME. The data points are shifted along the $x$ axis for clarity.}
	\label{fig-11}
\end{figure}

Figure~\ref{fig-11} shows the centrality dependence of the $b/a$ ratio, as predicted by the AMPT model with varying CME strengths. Focusing on the 20-50\% centrality bins in Au+Au collisions at $\sqrt{s_{_{\rm NN}}} = 200$ GeV, the $b/a$ ratio can be approximated by a constant function, yielding $b/a = 0.88( \pm 0.08)$. Note that our previous study shows that isobar collisions yield $b/a = 0.65( \pm 0.18)$~\cite{Chen:2023jhx}. This result implies that the relative ratio of CME signals across different planes does not simply invert the elliptic flow ratio among these planes. This finding has significant implications for extracting the fraction of the CME signal within the $\Delta\gamma$ observable, which will be discussed further.  

\begin{figure}
	\includegraphics[scale=0.40]{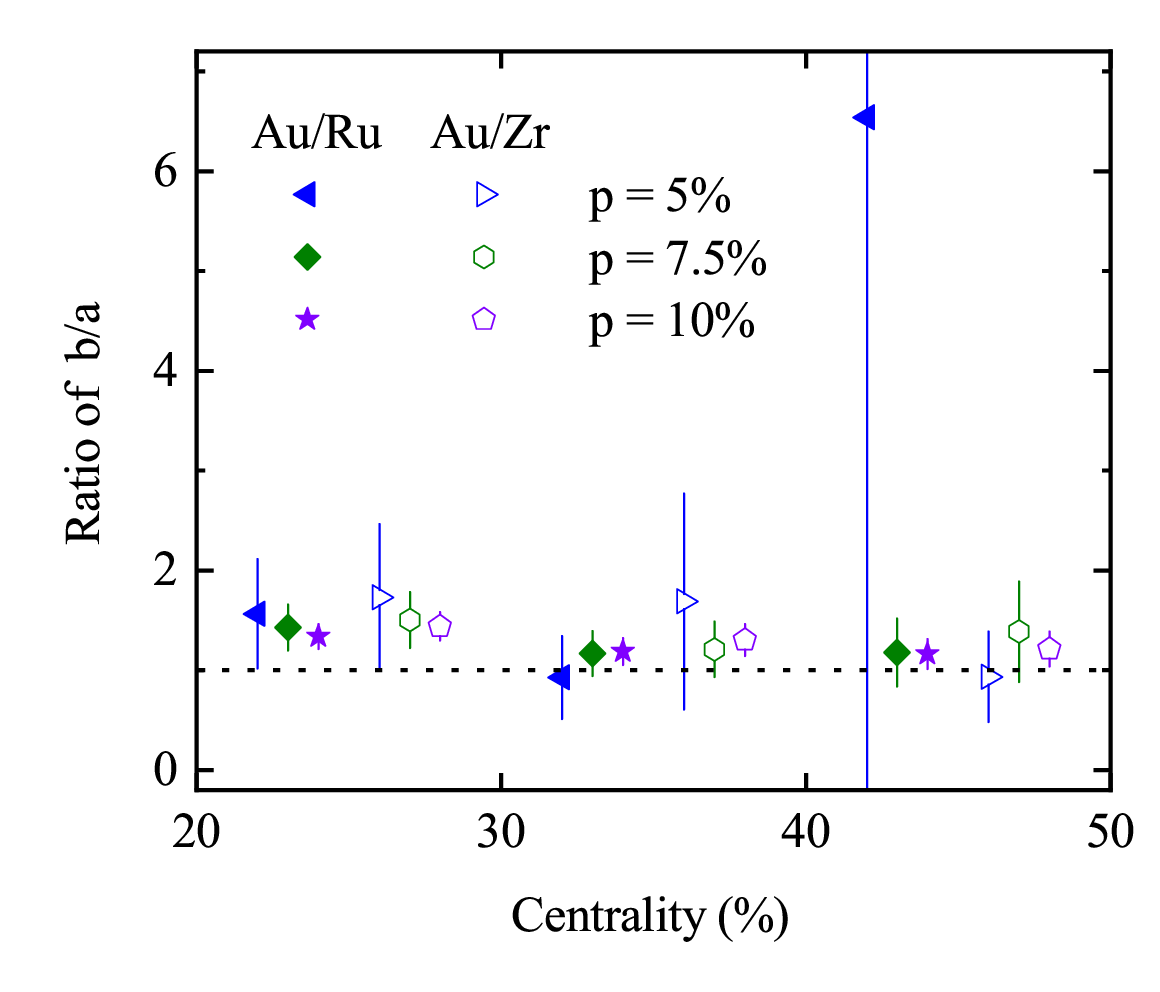}
	\caption{(Color online) The centrality dependence of the  $b/a$ ratios of Au+Au collisions to Ru+Ru collisions and Zr+Zr collisions, respectively, at $\sqrt{s_{_{\rm NN}}} = 200$ GeV from the AMPT model with different strengths of the CME. The data points are shifted along the $x$ axis for clarity.}
	\label{fig-12}
\end{figure}

Figure~\ref{fig-12} shows the centrality dependence of $b/a$ for Au+Au collisions relative to Ru+Ru and Zr+Zr collisions, respectively, based on the AMPT model with varying CME strengths. 
Since the value of \( a \) remains nearly constant, the ratio of \( b/a \) follows the trend of the ratio of \( b \) shown in Figure~\ref{fig-10}.

\begin{figure}
	\includegraphics[scale=0.4]{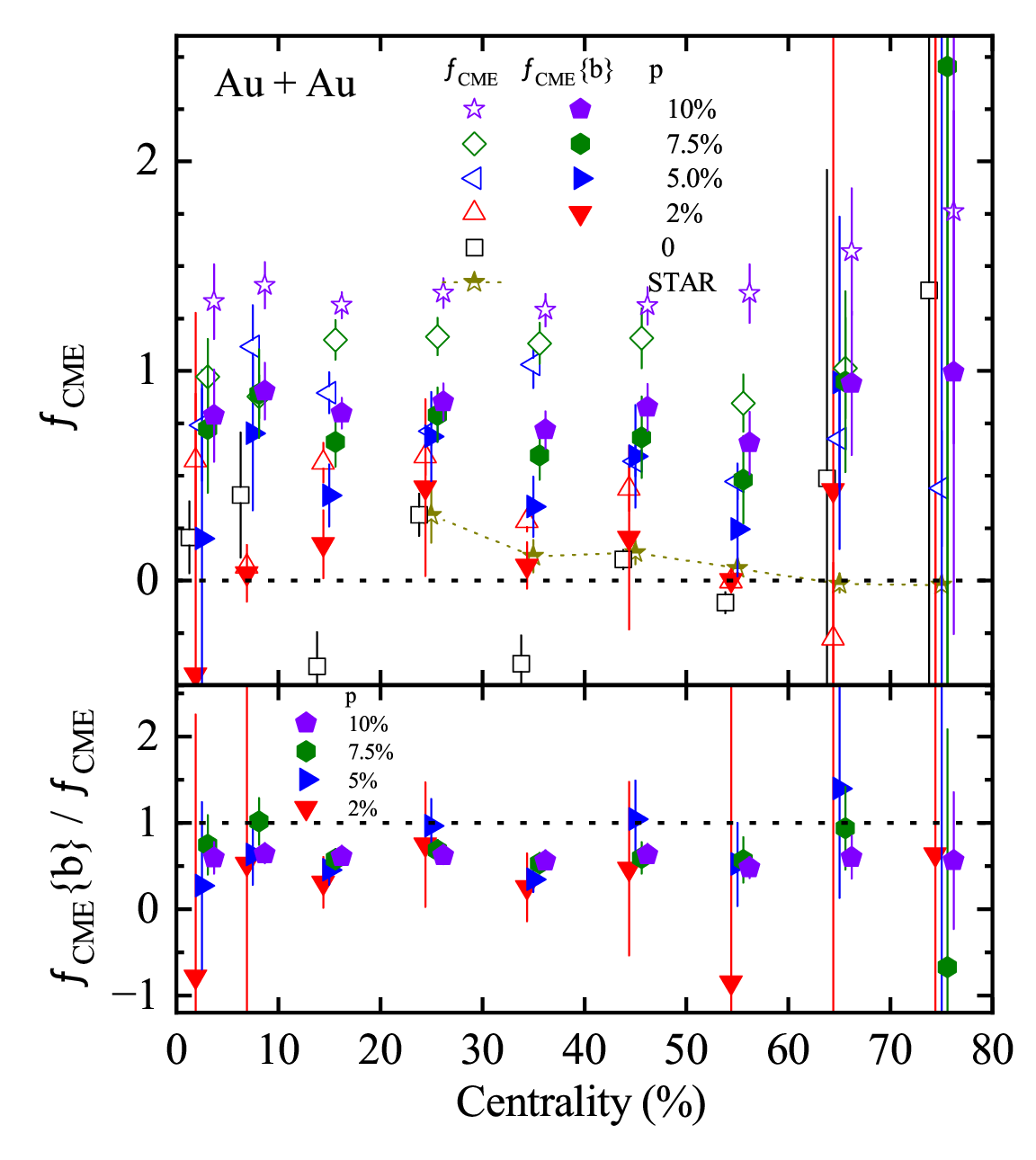}
	\caption{(Color online) Upper panel: The centrality dependence of $f_{\mathrm{CME}}\{b\}$ and $f_{\mathrm{CME}}$ in Au+Au collisions at $\sqrt{s_{_{\rm NN}}} = 200$ GeV from the AMPT model with different strengths of the CME, in comparison with the STAR data~\cite{STAR:2021pwb}. The solid and open symbols represent the results for $f_{\mathrm{CME}}\{b\}$ and $f_{\mathrm{CME}}$, respectively. Lower panel: The centrality dependence of the ratio of $f_{\mathrm{CME}}\{b\}$ to $f_{\mathrm{CME}}$. The data points are shifted along the $x$ axis for clarity.}
	\label{fig-13}
\end{figure}

The upper panel of Fig.~\ref{fig-13} shows the centrality dependence of two types of $f_{\mathrm{CME}}$ based on the AMPT model, with varying CME strengths. Open and solid symbols represent $f_{\mathrm{CME}}$ and $f_{\mathrm{CME}}\{b\}$ calculated using Eqs.~(\ref{equ.09}) and (\ref{equ.12}), respectively. Our results are consistent with the STAR experimental data~\cite{STAR:2021pwb}, favoring the cases of $p=0\%$ or $p=2\%$. Notably, for the 20-50\% centrality bins, when $p\neq0$, we find that $f_{\mathrm{CME}}\{b\}$ is smaller than $f_{\mathrm{CME}}$. The lower panel of Fig.~\ref{fig-13} shows the centrality dependence of the ratio $f_{\mathrm{CME}}\{b\}$ to $f_{\mathrm{CME}}$, which is less than unity for the 20-50\% centrality bins. This implies that assuming $b=a$ would lead to an overestimation of the CME signal fraction within the $\Delta\gamma$ observable.  

\begin{figure}
	\includegraphics[scale=0.4]{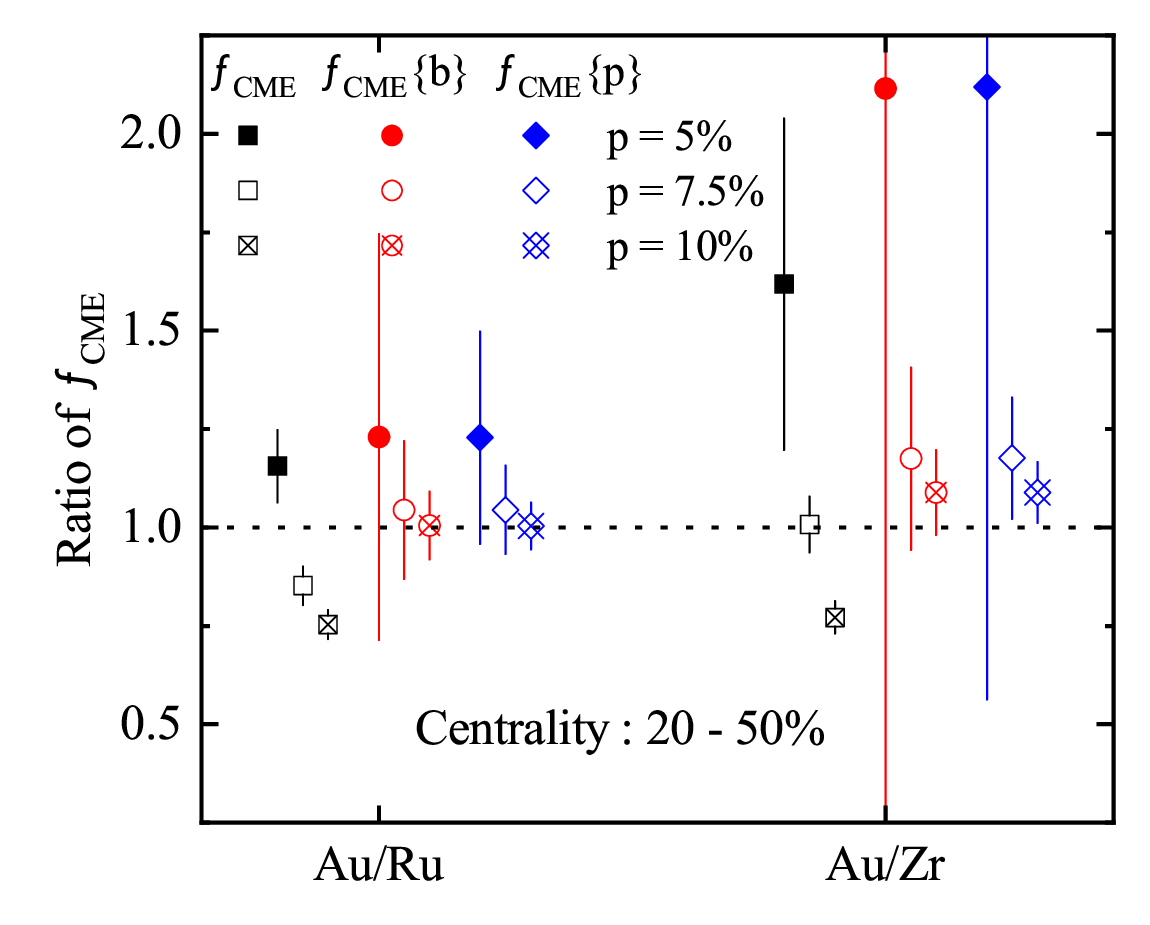}
	\caption{(Color online) The $f_{\mathrm{CME}}$,$f_{\mathrm{CME}}\{b\}$ and $f_{\mathrm{CME}}\{p\}$  ratios of Au+Au collisions to Ru+Ru collisions and Zr+Zr collisions，respectively, in 20-50\% centrality bins at $\sqrt{s_{_{\rm NN}}} = 200$ GeV from the AMPT model with different strengths of the CME. The solid, open symbols, and open symbols with X represent the results for the AMPT model with different strengths of the CME, respectively, in comparison with the STAR data~\cite{STAR:2021pwb,STAR:2021mii}. The data points are shifted along the $x$ axis for clarity.}
	\label{fig-14}
\end{figure}

Figure~\ref{fig-14} shows the $f_{\mathrm{CME}}$, $f_{\mathrm{CME}}\{b\}$, and $f_{\mathrm{CME}}\{p\}$ ratios for Au+Au collisions relative to Ru+Ru and Zr+Zr collisions, respectively, in the 20-50\% centrality bins at $\sqrt{s_{_{\rm NN}}} = 200$ GeV. Note that the STAR data for Ru+Ru and Zr+Zr collisions are obtained by averaging the data reported in Ref~\cite{STAR:2021mii}, which indicates that without accounting for the effect of \( b \), the fraction of CME signals in \( \Delta\gamma \) is slightly larger in isobar collisions than in Au+Au collisions.
The $f_{\mathrm{CME}}\{p\}$ ratio is calculated following the method described in Ref.~\cite{Chen:2023jhx}. However, the ratio for Au+Au collisions relative to isobar collisions increases after applying the correction for $b$, 
which indicates that after accounting for \( b \), the fraction of CME signals in \( \Delta\gamma \) is greater for Au+Au collisions than for isobar collisions. Therefore, it is important to take the effect of \( b \) into account to achieve the real fact about the CME.

As the introduced CME signal strength increases, the ratios calculated by all three methods decrease. This is because the $A/a$ value increases more significantly in isobar collisions than in Au+Au collisions, leading to a reduction in the calculated ratios. It is also observed that the Au/Ru ratio is smaller than the Au/Zr ratio in the presence of CME signals, indicating that the fraction of CME signal is higher in Ru+Ru than in Zr+Zr.  

\begin{figure}
	\includegraphics[scale=0.4]{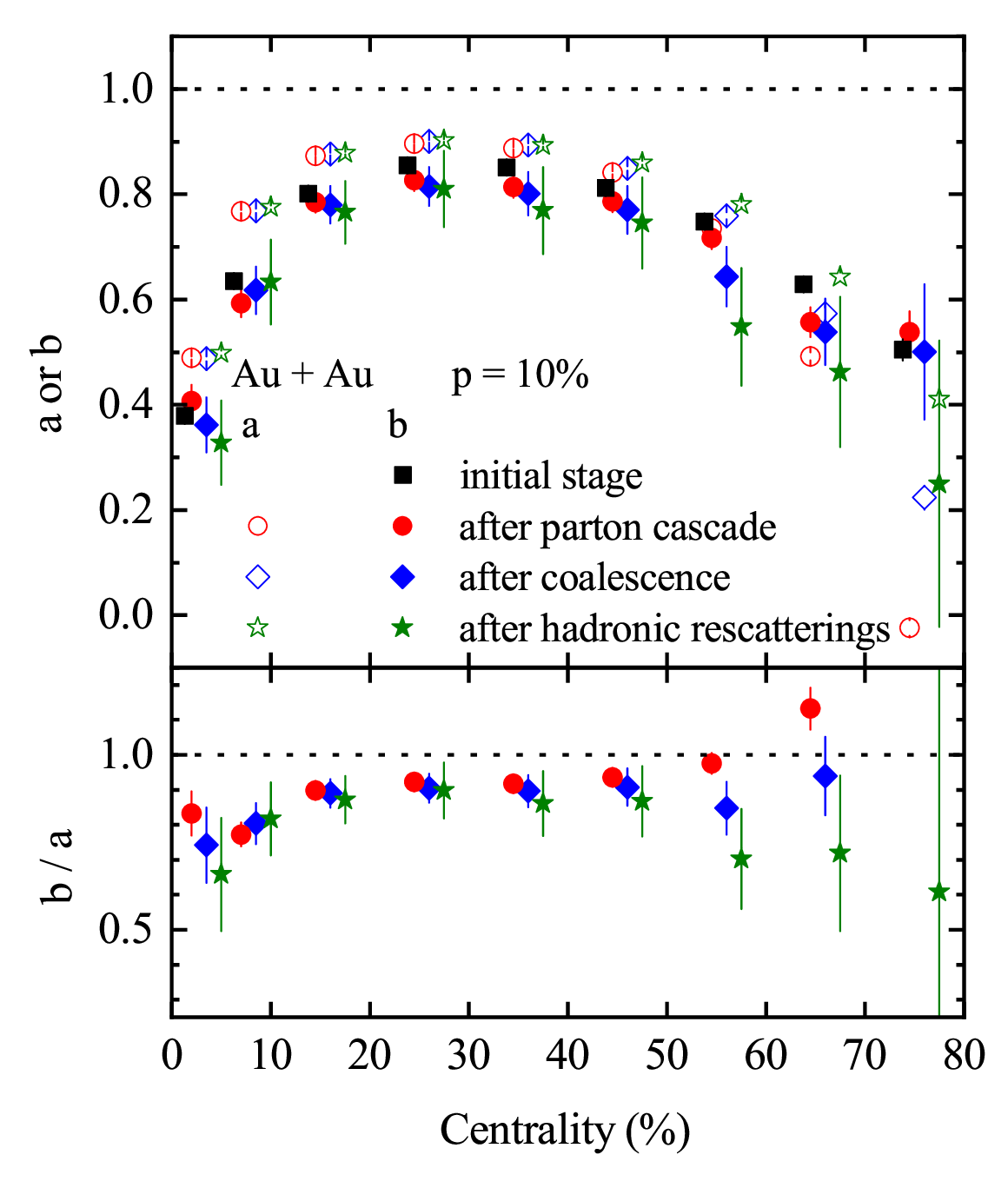}
	\caption{(Color online) Upper panel: The centrality dependence of \( a \) and \( b \) in Au+Au collisions at \(\sqrt{s_{_{\rm NN}}} = 200\) GeV for four different stages of the AMPT model with a CME strength of \( p = 10\%\). Open and solid symbols represent the results for \( a \) and \( b \), respectively. Lower panel: The centrality dependence of the ratio \( b/a \) for the different stages in Au+Au collisions. The same symbols as \( b \) in the upper panel are used to indicate different stages. The data points are shifted along the $x$ axis for clarity.
	}
	\label{fig-15}
\end{figure}

The preceding results demonstrate that the ratio $b/a = 0.88( \pm 0.08)$ significantly influences the final result of $f_{\mathrm{CME}}$ in the 20-50\% centrality bins for Au+Au collisions. To elucidate the origin of this relation, we investigate the evolution of $a$ and $b$ at different stages of Au+Au collisions at $\sqrt{s_{_{\rm NN}}} = 200$ GeV in the AMPT model, assuming a CME strength of $p=10$\%. We focus on four distinct stages: the initial stage, after the parton cascade, after coalescence, and after hadron rescatterings. As shown in the upper panel of Fig.~\ref{fig-15}, the value of $a$ remains constant during the last three stages. The initial stage is excluded from the $a$ calculation, as the elliptic flow is initially zero. In contrast, the value of $b$ is consistently smaller than $a$ and decreases with each stage in the 20-50\% centrality bins. The lower panel of Fig.~\ref{fig-15} illustrates the decreasing trend of the $b/a$ ratio during stage evolution, primarily driven by the decrease in $b$. This decrease in $b$ suggests that the correlation between CME signals across different planes becomes increasingly decorrelated, which can be attributed to the effects of final-state interactions during the evolution of heavy-ion collisions~\cite{Ma:2011uma,Huang:2019vfy,Huang:2022fgq,Zhao:2022grq}.  

\begin{figure}
	\includegraphics[scale=0.4]{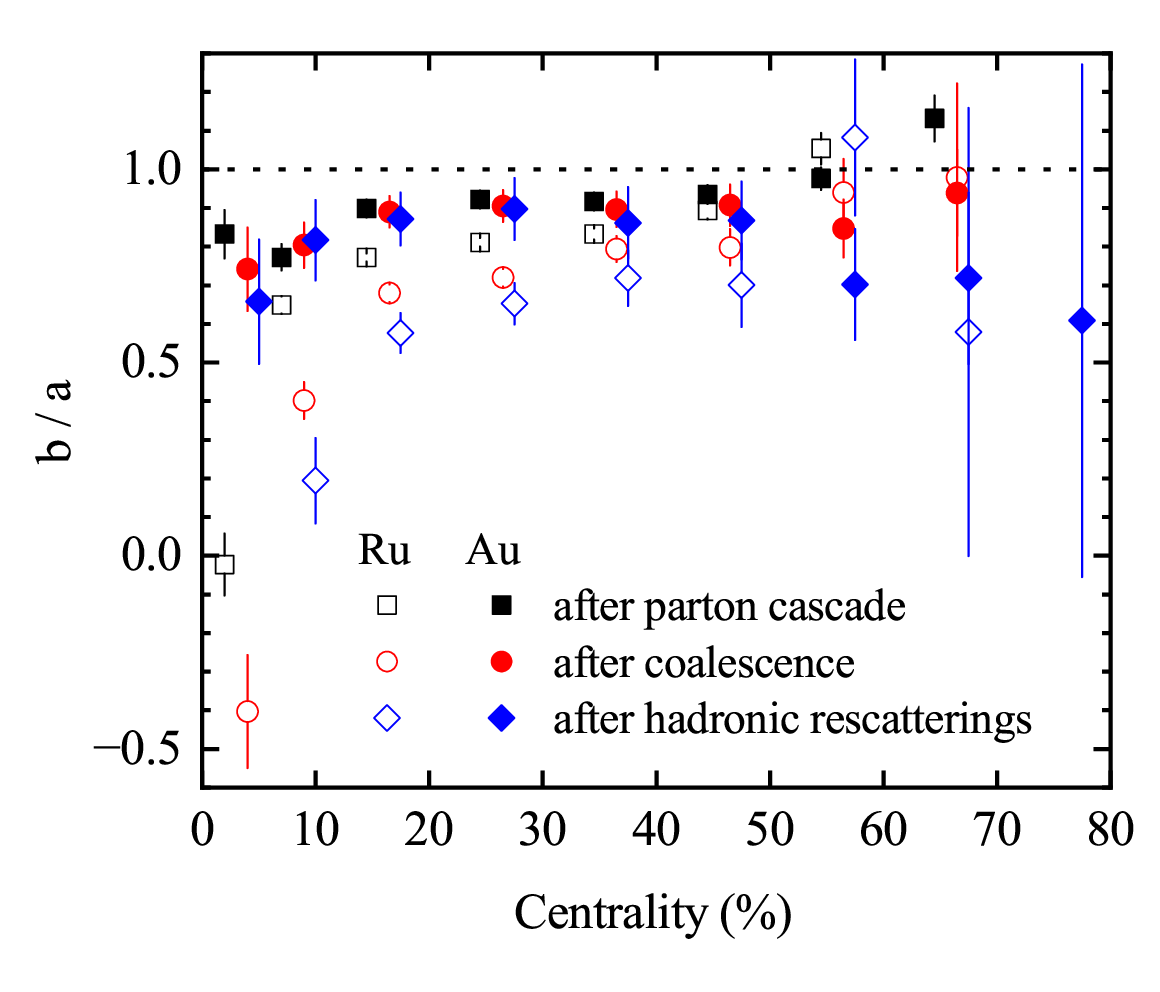}
	\caption{(Color online) The centrality dependence of
    the ratio of $b/a$  in Ru+Ru~\cite{Chen:2023jhx} and Au+Au collisions at $\sqrt{s_{_{\rm NN}}} = 200$ GeV for different stages from the AMPT model with the CME strength of $p=10$\%. The open and solid symbols represent the results for Ru+Ru and Au+Au, respectively. The data points are shifted along the $x$ axis for clarity.}
	\label{fig-16}
\end{figure}

In Ref.~\cite{Ma:2011uma}, the authors demonstrated that final-state interactions in relativistic heavy-ion collisions significantly suppress the initial charge separation, with a reduction factor reaching up to an order of magnitude. In our previous work~\cite{Chen:2023jhx}, we found that due to the anisotropic overlap zone, these interactions not only reduce the magnitude of the CME current but also alter its direction, resulting in a modified maximum current orientation. Fig.~\ref{fig-16} shows the centrality dependence of the $b/a$ ratio in Ru+Ru~\cite{Chen:2023jhx} and Au+Au collisions at $\sqrt{s_{_{\rm NN}}} = 200$ GeV for different stages in the AMPT model with a CME strength of $p=10$\%. For the 20-50\% centrality bins, the $b/a$ ratios show a decreasing trend with stage evolution for both Ru+Ru and Au+Au collisions. Furthermore, the $b/a$ values for Ru+Ru collisions are consistently smaller than those for Au+Au collisions across different stages and decrease more rapidly. This observation implies that the effect of final-state interactions on the decorrelation of the CME signals relative to the spectator and participant planes are less pronounced in Au+Au collisions, making it experimentally more favorable to detect the CME signal using the two-plane method. In our previous AMPT study~\cite{Ma:2011uma,Chen:2023jhx}, it was shown that assuming \( b = a \) in isobar collisions overestimates the final-state CME fraction in the measured \( \Delta\gamma \) observable, since the signal is subsequently damped and washed out during the final-state evolution of relativistic heavy-ion collisions. We find that \( b/a = 0.88 \pm 0.08 \) is larger in Au+Au collisions than in isobar collisions at the same centrality (\( b/a = 0.65 \pm 0.18 \)), indicating that the signal is damped and washed out to a lesser extent in Au+Au collisions than in isobar collisions.

\begin{figure}
	\includegraphics[scale=0.8]{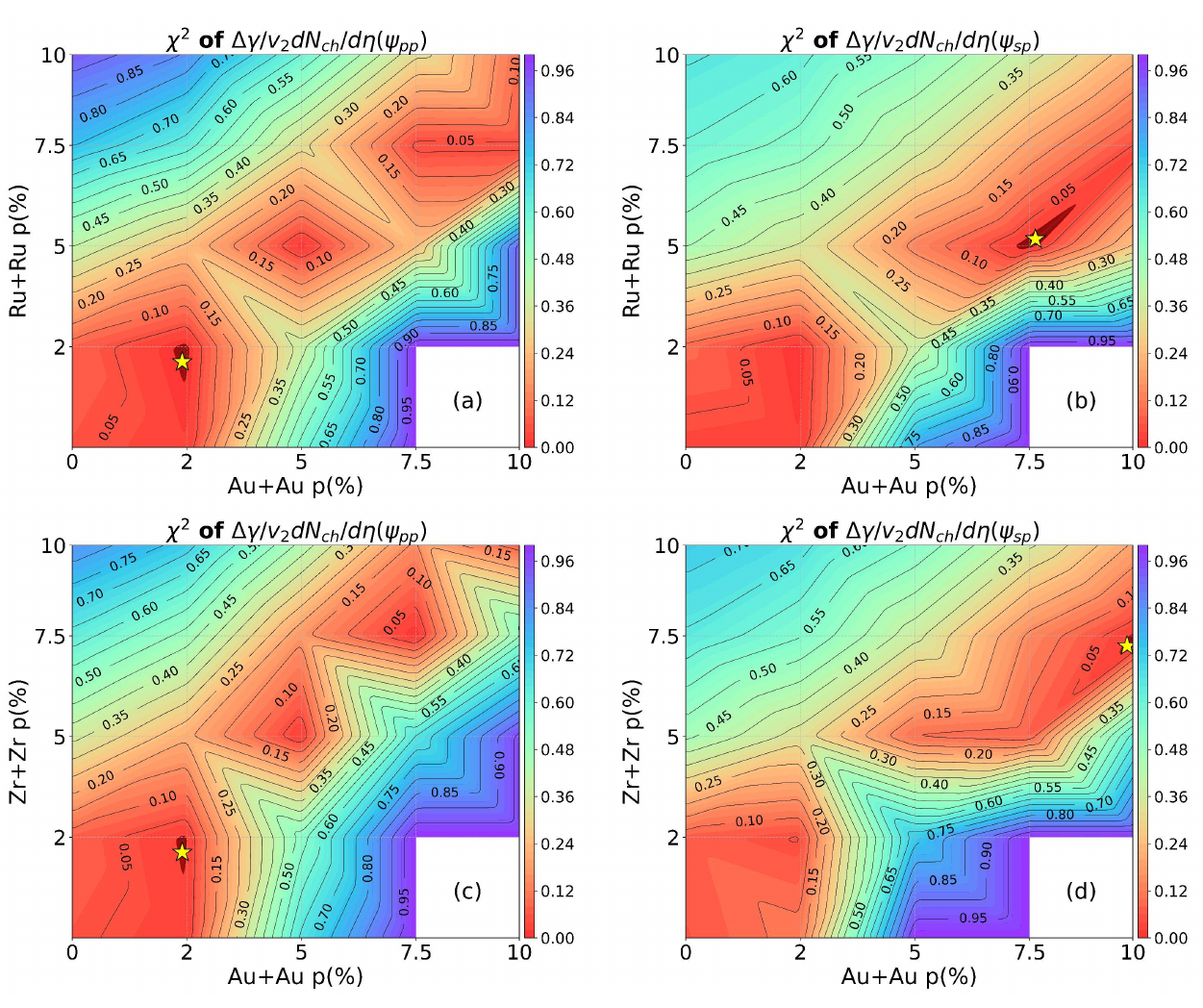}
	\caption{
    (Color online) The two-dimensional normalized chi-square distribution with respect to the different CME signal strengths in Au+Au collisions and the different CME signal strengths in isobar collisions. The chi-square ($\chi^2$) value represents the goodness of how well our model describes the experimental data for the CME observable of $\Delta \gamma / v_2 d N_{ch} / d \eta$. The panels (a) and (b) show the results between Au+Au collisions and Ru+Ru collisions with respect to $\psi_{\rm PP}$ and $\psi_{\rm SP}$, respectively.  The panels (c) and (d) show the results between Au+Au collisions and Zr+Zr collisions with respect to $\psi_{\rm PP}$ and $\psi_{\rm SP}$, respectively. Note that the magnitudes of the normalized chi-square value vary with color, where the dark red region, highlighted as a pentagram, indicates the smallest $\chi^2$ value which corresponds to the best description. 
	}
	\label{fig-17}
\end{figure}

To constrain the CME strengths across Au+Au collisions and isobar collisions simultaneously, we performed a chi-square analysis using the following method. We aim to compare the CME observable between our results with different CME strengths and the experimental data for those three collision systems simultaneously. The chi-square for a CME observable $O$ in a centrality bin $i$ is defined as:
\begin{equation}\label{equ.14}
	\chi^2=\sum_{k=0}^n \frac{\left(O_i-E_i\right)^2}{w_o^2}.
\end{equation}
where we choose the CME observable of $O_i$ as a double ratio between Au+Au collisions and isobar collisions to reduce the effect of the background,
\begin{equation}\label{equ.14}
	O_i=\frac{\frac{\{\Delta \gamma(\psi)\}_\mathrm{Au}}{\{v_2(\psi)\}_\mathrm{Au}} \times \{d N_{c h} / d \eta\}_\mathrm{Au}}{\frac{\{\Delta \gamma(\psi)\}_\mathrm{Isobar}}{\{v_2(\psi)\}_\mathrm{Isobar}} \times \{d N_{c h} / d \eta\}_\mathrm{Isobar}},
\end{equation}
where $\psi$ is $\psi_{\rm PP}$ or $\psi_{\rm SP}$, respectively, and $d N_{c h} / d \eta$ denotes the number of charged particles. The $E_i$ represents the experimental results. The $w_o^2$ represents the error of the calculation results of $O_i$, and the $i$ denotes in the 20-50\% centrality bin.  
Figure~\ref{fig-17} shows the results of the two-dimensional normalized chi-square distribution with respect to the different CME signal strengths in Au+Au collisions and the different CME signal strengths in isobar collisions. For the PP plane cases shown in panels (a) and (c), the AMPT simulation results align most closely with the experimental data when the CME strengths for Au+Au collisions, Ru+Ru and Zr+Zr collisions are all close to 2\%. This indicates that for the PP plane, the experimental results tend to exhibit small CME signals for all three collision systems, since the CME signal in the PP plane is relatively insensitive to the SP plane. On the other hand, for the SP plane cases shown in panels (b) and (d), the AMPT simulation results align most closely with the experimental data when the CME strength in Au+Au collisions is 7.5\% and that in Ru+Ru collisions is 5\%, or when the CME strength in Au+Au collisions is 10\% and that in Zr+Zr collisions is 7.5\%. This supports that the experimental results for the SP plane tend to exhibit larger CME signals than those for the PP plane for all three collision systems, with stronger signals observed in Au+Au collisions than those in isobar collisions,  as the CME signal exhibits greater sensitivity to the SP plane than to the PP plane.

\section{Summary}
\label{sec:summary}

Using a multiphase transport model with varying CME strengths, we have extended our two-plane method analysis from isobar collisions to Au+Au collisions, both at $\sqrt{s_{_{\rm NN}}} = 200$ GeV. Our previous isobar collision studies revealed a significant difference ($b/a$ = 0.65 $\pm$ 0.18 ) in the CME signal-to-background ratio between two planes, complicating the CME signal extraction in isobar collisions~\cite{Chen:2023jhx}. However, this current Au+Au analysis demonstrates a reduced difference ($b/a$ = 0.88 $\pm$ 0.08)  in the CME signal-to-background ratio between two planes, which enhances the experimental reliability of two-plane method measurement in Au+Au collisions. Through comprehensive chi-square analysis across the three collision systems, we establish that Au+Au collisions exhibit stronger CME signature compared to the isobar systems, particularly when analyzed with respect to the spectator plane. These findings not only validate the improved applicability of the two-plane method in Au+Au collisions but also provide critical insights into the system size dependence of the CME observable, advancing our future experimental measurements of the chiral magnetic effect.

\section*{ACKNOWLEDGMENTS}
We thank Profs. Fuqiang Wang and Jie Zhao for their helpful discussions. This work is supported by the National Natural Science Foundation of China under Grants No.12147101, No. 12325507, and No.12105054, the National Key Research and Development Program of China under Grant No. 2022YFA1604900, and the Guangdong Major Project of Basic and Applied Basic Research under Grant No. 2020B0301030008.

\bibliography{fcme} % Produces the bibliography via BibTeX.
 
\end{CJK*}
\end{document}